\documentclass[11pt,oneside,onecolumn,draftclsnofoot,journal,a4paper]{IEEEtran}
\usepackage{bbm}
\usepackage{enumerate}
\usepackage{array}
\usepackage{dcolumn}
\usepackage{psfig}
\usepackage[intlimits]{amsmath}
\usepackage[center]{subfigure}
\usepackage{amsmath}
\usepackage{amssymb}
\usepackage{epsfig}
\usepackage{color,soul}

\newlength {\parwidth}
\newtheorem{thm}{Theorem}
\newtheorem{lem}{Lemma}

\newcounter{letter}
\newcommand{\eqabcbegin}{\setcounter{letter}{1}
            \renewcommand{\theequation}{\arabic{equation}\alph{letter}}}
\newcommand{\nexteqabc}{\addtocounter{letter}{1}\addtocounter{equation}{-1}}
\newcommand{\eqabcend}{\renewcommand{\theequation}{\arabic{equation}}}

\newcommand{\postscript}[4]
{
 \begin{figure}[htb]
 \par
 \hbox to 9cm {\vbox to #1
        {
        \vfill
        \includegraphics{#2.eps}
        }
      }
 \caption{#3}
 \label{#4}
 \end{figure}
}

\newcommand {\myvec}[1] {{\mbox{\boldmath $#1$}}}
\newcommand {\mymat}[1]  {{\mbox{\boldmath $#1$}}}

\def\Pr{\mathop{\mathrm{ Pr}}\nolimits}

\def\comment#1{}

\def\xs{\mbox{\begin{scriptsize}\boldmath{$x$}\end{scriptsize}}}
\def\ys{\mbox{\begin{scriptsize}\boldmath{$y$}\end{scriptsize}}}
\def\uus{\mbox{\begin{scriptsize}\boldmath{$u$}\end{scriptsize}}}

\newcommand {\defeq}{\stackrel{\triangle}{=}}

\newcommand {\uh}   {\myvec {h}}

\newcommand {\ua}   {\myvec {a}}

\newcommand {\hub}   {\hat{\myvec {b}}}
\newcommand {\huB}   {\hat{\myvec {B}}}
\newcommand {\buB}   {\bar{\myvec {B}}}

\newcommand {\uu}   {\myvec {u}}
\newcommand {\ui}   {\myvec {i}}
\newcommand {\uj}   {\myvec {j}}
\newcommand {\un}   {\myvec {n}}
\newcommand {\um}   {\myvec {m}}

\newcommand {\uo}   {\myvec {0}}

\newcommand {\us}   {\myvec {s}}
\newcommand {\uv}   {\myvec {v}}
\newcommand {\up}   {\myvec {p}}
\newcommand {\tup}   {\tilde {\up}}
\newcommand {\tp}   {\tilde {p}}

\newcommand {\ux}   {\myvec {x}}
\newcommand {\uy}   {\myvec {y}}
\newcommand {\uw}   {\myvec {w}}
\newcommand {\A}    {\mymat {A}}
\newcommand {\B}    {\mymat {B}}

\newcommand {\G}    {\mymat {G}}
\newcommand {\F}    {\mymat {F}}
\newcommand {\uL}    {\mymat {L}}
\newcommand {\uU}    {\mymat {U}}

\newcommand {\hB}    {\mymat {\hat{B}}}

\newcommand {\uE}    {\mymat {{E}}}

\newcommand {\uPi}    {\mymat {\Pi}}
\newcommand {\uLam}    {\mymat {\Lambda}}
\newcommand {\I}    {\mymat {I}}
\renewcommand {\L}  {\mymat {\Lambda}}

\newcommand {\D}    {\mymat {D}}
\renewcommand {\P}    {\mymat {P}}
\newcommand {\tuP}    {\widetilde {\P}}
\newcommand {\tP}    {\widetilde {P}}

\newcommand {\bbR}    {\mymat {\mathbb{R}}}
\newcommand {\bbC}    {\mymat {\mathbb{C}}}
\newcommand {\GFP}    {\mymat {\mathbb{GF}}(P)}
\newcommand {\GFii}    {\mymat {\mathbb{GF}}(2)}
\newcommand {\GFiii}    {\mymat {\mathbb{GF}}(3)}

\newcommand {\V}    {\mymat {V}}

\newcommand {\hp}   {\hat{p}}
\newcommand {\bPPx}   {\mymat{\mathcal{P}}_{\xs}}
\newcommand {\bPPy}   {\mymat{\mathcal{P}}_{\ys}}
\newcommand {\bPPu}   {\mymat{\mathcal{P}}_{\uus}}
\newcommand {\PPu}   {\mathcal{P}_{\uus}}
\newcommand {\PPx}   {\mathcal{P}_{\xs}}
\newcommand {\tbPPu}   {\widetilde{\mymat{\mathcal{P}}}_{\uus}}
\newcommand {\tPPu}   {\widetilde{\mathcal{P}}_{\uus}}
\newcommand {\hbPPx}   {\widehat{\mymat{\mathcal{P}}}_{\xs}}
\newcommand {\hbPPy}   {\widehat{\mymat{\mathcal{P}}}_{\ys}}
\newcommand {\htPPx}   {\widehat{\widetilde{\mathcal{P}}}_{\xs}}
\newcommand {\htbPPx}   {\widehat{\widetilde{\mymat{\mathcal{P}}}}_{\xs}}

\newcommand {\hPPx}   {\widehat{\mathcal{P}}_{\xs}}
\newcommand {\tPPx}   {\widetilde{\mathcal{P}}_{\xs}}

\newcommand {\tbPPx}   {\widetilde{\mymat{\mathcal{P}}}_{\ux}}

\newcommand {\bH}   {H_{mar}}

\newcommand {\buy}   {\bar{\uy}}

\newcommand {\ual}   {\myvec {\alpha}}

\newcommand {\hPr}   {\widehat{\rm{Pr}}}

\begin{document}

\title{Independent Component Analysis Over Galois Fields}
\author{Arie Yeredor, \em{Senior Member, IEEE}%
\thanks{The author is with the Department of Electrical
Engineering - Systems, Tel-Aviv University, Tel-Aviv, Israel. Some
part of this work was presented at ICA'07 \cite{ICA07}. }}

\maketitle
\begin{abstract}

We consider the framework of Independent Component Analysis (ICA)
for the case where the independent sources and their linear
mixtures all reside in a Galois field of prime order $P$.
Similarities and differences from the classical ICA framework
(over the Real field) are explored. We show that a necessary and
sufficient identifiability condition is that none of the sources
should have a Uniform distribution. We also show that pairwise
independence of the mixtures implies their full mutual
independence (namely a non-mixing condition) in the binary ($P=2$)
and ternary ($P=3$) cases, but not necessarily in higher order
($P>3$) cases. We propose two different iterative separation (or
identification) algorithms: One is based on sequential
identification of the smallest-entropy linear combinations of the
mixtures, and is shown to be equivariant with respect to the
mixing matrix; The other is based on sequential minimization of
the pairwise mutual information measures. We provide some basic
performance analysis for the binary ($P=2$) case, supplemented by
simulation results for higher orders, demonstrating advantages and
disadvantages of the proposed separation approaches.

\end{abstract}

\section{Introduction}

Independent Component Analysis (ICA, see, e.g.,
\cite{comon,cardoso,ICAbook} for some of the fundamental
principles) addresses the recovery of unobserved, statistically
independent source signals from their observed linear (and
invertible) mixtures, without prior knowledge of the mixing matrix
or of the sources' statistics. Classically, the ICA framework
assumes that the sources and the mixing (hence, also the
observations) are defined over the field of real-valued numbers
$\bbR$, with some exceptions (e.g.,
\cite{jan}) that assume the field of complex-valued numbers
$\mathbb{C}$. It might be interesting, though, at least from a
theoretical point of view, to explore the applicability of ICA
principles in other algebraic fields.

In this work we consider ICA over Galois Fields of prime order
$P$, denoted $\GFP$, such that the sources and the mixing-matrix'
elements can all take only a finite number of values, defined by
the set $\{0,1,...,P-1\}$ (or by some offset, isomorphic version
thereof), and where addition and multiplication are applied modulu
$P$, thereby returning values in the same set.

For example, in the field $\GFii$ of binary numbers $\{0,1\}$,
addition is obviously equivalent to the ``Exclusive Or" (XOR)
operation, denoted $z=x\oplus y$ (where $z$ equals $1$ if $x\ne y$
and equals $0$ otherwise). Multiplication (either by $0$ or by
$1$) is defined in the ``usual" way in this case.

In the field $\GFiii$ of ternary numbers $\{0,1,2\}$, where
addition and multiplication are defined modulu $3$ (similarly
denoted $z=x\oplus y$), it is sometimes more convenient to
consider the offset group $\{0,1,-1\}$. In this group,
multiplication can still be defined in the ``usual" way, since
ordinary multiplication of any two numbers in this group returns a
number in the group. Obviously, the two sets $\{0,1,2\}$ and
$\{0,1,-1\}$ are isomorphic in $\GFiii$, and will be used
interchangeably in the sequel.

A fundamental difference, at least in the context of ICA, between
random variables over $\bbR$ and over $\GFP$ is the following: Let
$u$ and $v$ be two statistically independent, non-degenerate
(namely, non-deterministic) random variables, and consider the
random variable $w$, given by any non-trivial linear combination
of $u$ and $v$. In $\bbR$, $v$ and $w$ cannot be statistically
independent (they are obviously correlated), no matter how $u$ and
$v$ are distributed. However, as we shall show in Section
\ref{sec:form}, in $\GFP$ $v$ and $w$ may indeed be statistically
independent, and this happens if and only if the distribution of
$u$ is uniform (taking each of the $P$ values with equal
probabilities).

In a sense, this property tags the uniform distribution as the
``problematic" distribution in ICA over $\GFP$, reminiscent of the
role taken by the Gaussian distribution in ICA over $\bbR$. Note
that these two distributions share additional related properties
in their respective fields: They are both (under mild regularity
conditions) limit-distributions of an infinite sum of independent
random variables; and they are both ``maximum entropy"
distributions (subject to a variance constraint for the Gaussian
distribution in $\bbR$). So, loosely stated, in the same way that
a linear combination of independent random variables over $\bbR$
tends to be ``more Gaussian", a linear combination of independent
random variables over $\GFP$ tends to be ``more uniform".

Nevertheless, there still remain some essential differences
between the roles of these distributions in the respective
contexts. For example, in $\GFP$, if (at least) one of random
variables in the linear combination of independent variables is
uniform, the resulting distribution would be exactly uniform as
well, no matter how the other random variables are distributed.
Evidently, this property does not hold for Gaussian distributions
over $\bbR$.

Therefore, as we shall show, these properties lead to an
identifiability condition for ICA over $\GFP$, which is
reminiscent of, but certainly not equivalent to, a well-known
identifiability condition over $\bbR$. More specifically, the
identifiability condition for ICA over $\bbR$ requires that {\em
not more than one} of the sources be Gaussian. Our identifiability
condition for ICA over $\GFP$ requires that {\em none} of the
sources be uniform. The key to this identifiability condition is
the property that the entropy of any linear combination of
statistically independent random variables over $\GFP$ is larger
than the entropy of the largest-entropy component, as long as this
component is not uniform. Therefore, if none of the sources is
uniform, then, at least conceptually, a possible separation
approach is to look for the (inverse) linear transformation, which
minimizes the empirical marginal entropies of the resulting linear
combinations. However, since an exhaustive search for this
transformation would often be prohibitively computationally
expensive, we shall propose an alternative, computationally
cheaper method for entropy-based identification.

Another possible, somewhat different separation approach is the
following. One of the key observations in ICA over $\bbR$ is that,
under the identifiability condition and due to the
Darmois-Skitovitch theorem (e.g., \cite{DST}, p.218),
pairwise-independence of the mixtures implies their full mutual
independence, which in turn implies a non-mixing condition
(namely, separation). Interestingly, we shall show that our
general identifiability condition is necessary and sufficient to
guarantee a similar property for ICA over $\GFii$ and $\GFiii$,
but is generally insufficient for this property to hold in $\GFP$
for $P>3$. Thus, another possible identification approach (in
$\GFii$ and in $\GFiii$ only) is to look for an invertible linear
transformation of the observations, which makes the resulting
signals ``as empirically pairwise-independent as possible" - a
property which is easier to quantify and measure than full
independence (being quadratic, rather than exponential, in the
number of sources $K$). Again - since an exhaustive search is
often not feasible, we shall propose a different, sequential
method for this approach.


A common assumption in the design and analysis of classical ICA
methods over $\bbR$, is that each of the sources has an
independent, identically distributed (iid) time-structure. Our
discussion in this paper would be similarly restricted along the
same line. We note, however, that in equivalence to methods which
exploit possibly different temporal structures (e.g., spectral
diversity \cite{sobi}, non-stationarity \cite{nonstat}, etc.) over
$\bbR$, similar extensions of our results would be possible in
similar cases over $\GFP$. However, we defer the exploration of
such cases to future work.

The paper is structured as follows. In the next section we review
some fundamental properties of random variables and random vectors
in $\GFP$, which will be useful in subsequent derivations. In
Section \ref{sec:form} we outline the problem formulation and
present our general identifiability condition. In Section
\ref{sec:pair} we explore the relation between pairwise
independence and full independence, showing that in an invertible
linear mixture, the former implies the latter in $\GFii$ and in
$\GFiii$, but not necessarily in Galois fields of higher orders.
We then proceed to propose two different separation algorithm in
Section \ref{sec:algos}. A rudimentary performance analysis for
the simple binary case ($P=2$) is provided in Section
\ref{sec:analsim}, supplemented with supporting simulation results
which extend to larger-scale scenarios. Our work is summarized
with concluding remarks in Section
\ref{sec:conc}.

We shall denote addition, subtraction and multiplication over
$\GFP$ (namely, modulu $P$) by $\oplus$, $\ominus$ and $\otimes$,
respectively, with multiplication preceding addition and
subtraction in the order of operations. Vector multiplication will
be denoted by $\circ$, such that if $\ua=[a_1\;\cdots\;a_K]^T$ and
$\ux=[x_1\;\cdots\;x_K]^T$, then
\begin{equation}
\ua^T\circ\ux=a_1\otimes x_1\oplus a_2\otimes x_2\oplus\cdots
\oplus a_K\otimes x_K.
\end{equation}
Similarly, if $\A$ is an $L\times K$ matrix in $\GFP$, its product
with $\ux$ is denoted $\A\circ\ux$, an $L\times 1$ vector whose
elements are the products of the respective rows of $\A$ with
$\ux$.

\section{Characterization of random variables and random vectors in $\GFP$}
\label{sec:char}

We begin by briefly outlining some of the basic essential
properties and definitions of our notations for random variables
and random vectors in $\GFP$, which we shall use in the sequel.

A random variable $u$ in $\GFP$ is characterized by a discrete
probability distribution, fully described by a vector
$\up_u=[p_u(0)\;p_u(1)\;\cdots\;p_u(P-1)]^T\in\bbR^P$, whose
elements $p_u(m)$ are $\Pr\{u=m\}$, the probabilities of $u$
taking the values $m\in\{0,\ldots,P-1\}$. Evidently, all the
elements of $\up_u$ are non-negative and their sum equals $1$. We
shall refer to $\up_u$ as the {\em probability vector} of $u$. The
{\em entropy} of $u$ is  given by
\begin{equation}
\label{H}
H(u)=-\sum_{m=0}^{P-1}p_u(m)\log p_u(m).
\end{equation}
By maximizing with respect to $\up_u$, it is easy to show that
among all random variables in $\GFP$, the uniform random variable
(taking all values in $\GFP$ with equal probability
$\tfrac{1}{P}$) has the largest entropy, given by $\log P$. Note
that it is convenient to use a base-$P$ logarithm $\log_P$ (rather
than the more commonly-used $\log_2$) in this context, such that
the entropies of all (scalar) random variables in $\GFP$ are
confined to $[0,1]$. Note, in addition, that since multiplication
by a constant over $\GFP$ is bijective, the entropy of a random
variable in $\GFP$ is invariant under such multiplication (which
merely re-arranges the terms in the sum
\eqref{H}).

The {\em characteristic vector} of $u$ is denoted
$\tup_u=[\tp_u(0)\;\tp_u(1)\;\cdots\;\tp_u(P-1)]^T\in\bbC^P$, and
its elements are given by the discrete Fourier transform (DFT) of
the elements of $\up$:
\begin{equation}
\tp_u(n)=E[W_P^{nu}]=\sum_{m=0}^{P-1}p_u(m)W_P^{mn} \;\;\; n=0,\ldots,P-1,
\end{equation}
where the ``twiddle factor" $W_P$ is defined as $W_P=e^{-j2\pi/P}$
(note that the modulu-$P$ operation is inherently present in the
exponential part, so $W_P^{mn}$ is equivalent to $W_P^{m\otimes
n}$). Like the probability vector $\up_u$, the characteristic
vector $\tup_u$ provides full statistical characterization of the
random variable $u$, since $\up_u$ can be directly obtained from
$\tup_u$ using the inverse DFT.

\noindent
The following basic properties of $\tup_u$ can be easily obtained:
\renewcommand{\theenumi}{P\arabic{enumi}}
\begin{enumerate}
\item $\tp_u(0)=1$; \label{p0}
\item Since $\up_u$ is real-valued, $\tp_u(n)=\tp_u^*(P-n)$ (where the superscript $^*$ denotes the
complex-conjugate);\label{Pconj}
\item $u$ is uniform (namely, $p_u(m)=\tfrac{1}{P}$ $\forall m$) $\Leftrightarrow$ $\tp_u(n)=0$ $\forall
n\ne0$; \label{Puni}
\item $u$ is degenerate (namely, $p_u(M)=1$ for some $M$) $\Leftrightarrow$
$\tp_u(n)=W_P^{nM}$ $\forall n$; \label{Pdeg}
\item $|\tp_u(n)|\le 1$ $\forall n$, where for $n\ne
0$ equality holds if and only if (iff) $u$ is degenerate.
\label{Pabs}
\end{enumerate}
\renewcommand{\theenumi}{\arabic{enumi}}

\noindent
Note that in the particular cases of $\GFii$ and $\GFiii$ we have
the following simplifications:
\begin{itemize}
\item In $\GFii$, the only free parameter in $\tup_u\in\bbR^2$ is
$\tp_u(1)$, to which we shall refer as
\begin{equation}
\theta_u\defeq\tp_u(1)=p_u(0)-p_u(1)=1-2p_u(1).
\end{equation}
Thus $\tup_u=[1\;\theta_u]^T$;
\item In $\GFiii$, there is also a single (yet complex-valued)
free parameter in $\tup_u\in\bbC^3$, to which we shall refer as
\begin{equation}
\label{xiu}
\xi_u\defeq\tp_u(1)=p_u(0)+p_u(1)W_3^{-1}+p_u(2)W_3^{-2}=1-\tfrac{3}{2}(p_u(1)+p_u(2))+j\tfrac{\sqrt{3}}{2}(p_u(2)-p_u(1)).
\end{equation}
Thus $\tup_u=[1\;\xi_u\;\xi_u^*]^T$.
\end{itemize}
Note also that $\theta_u=E[W_2^{u}]=E[(-1)^u]$ and
$\xi_u=E[W_3^{u}]$.

For two random variables $u$ and $v$ in $\GFP$, the joint
statistics are completely described by the {\em joint
probabilities matrix} $\P_{u,v}\in\bbR^{P\times P}$, whose
elements are $P_{u,v}(m,n)=\Pr\{u=m,v=n\}$,
$m,n\in\{0,\ldots,P-1\}$. The joint entropy of $u$ and $v$ is
given by
\begin{equation}
H(u,v)=-\sum_{m,n=0}^{P-1}P_{u,v}(m,n)\log P_{u,v}(m,n).
\end{equation}
The random variables $u$ and $v$ are said to be {\em statistically
independent} iff $\P_{u,v}=\up_u\up_v^T$. By Jensen's inequality,
$H(u,v)$ satisfies $H(u,v)\le H(u)+H(v)$, with equality iff $u$
and $v$ are statistically independent. The {\em mutual
information} between $u$ and $v$ is the difference $I(u,v)\defeq
H(u)+H(v)-H(u,v)$, which is also the (non-negative)
Kullback-Leibler divergence between their joint distribution and
the product of their marginal distributions. The smaller their
mutual information, the ``more statistically independent" $u$ and
$v$ are; $I(u,v)$ vanishes if and only if $u$ and $v$ are
statistically independent.

The conditional distribution of $u$ given $v$ is given by
$\P_{u|v}\in\bbR^{P\times P}$ with elements
$P_{u|v}(m,n)=P_{u,v}(m,n)/p_v(n)=\Pr\{u=m|v=n\}$,
$m,n=0,\ldots,P-1$. The conditional entropy is defined as
\begin{equation}
H(u|v)=-\sum_{n=0}^{P-1}p_v(n)\sum_{m=1}^{P-1}P_{u|v}(m,n)\log
P_{u|v}(m,n),
\end{equation}
which can be easily shown to satisfy $H(u|v)=H(u,v)-H(v)$.

The joint {\em characteristic matrix} of $u$ and $v$, denoted
$\tuP_{u,v}\in\bbC^{P\times P}$, is given by the two-dimensional
DFT (2DFT) of $\P_{u,v}$,
\begin{equation}
\tP_{u,v}(m,n)=E[W_P^{mu+nv}]=\sum_{k,\ell=0}^{P-1}P_{u,v}(k,\ell)W_P^{\;mk+n\ell},
\end{equation}
and provides an alternative full statistical characterization of
$u$ and $v$. In particular, it is straightforward to show that
$\tuP_{u,v}$ satisfies $\tuP_{u,v}=\tup_u\tup_v^T$ iff $u$ and $v$
are statistically independent.

For a $K\times 1$ random vector $\uu$ whose elements
$u_1,\ldots,u_K$ are random variables in $\GFP$, the joint
statistics are fully characterized by the $K$-way {\em
probabilities tensor} $\bPPu\in\bbR^{P^{(\times K)}}$, whose
elements are the probabilities
$\PPu(m_1,\ldots,m_K)=\Pr\{u_1=m_1,\ldots,u_K=m_K\}$,
$m_1,\ldots,m_K\in\{0,\ldots,P-1\}$. Using vector-index notations,
where $\um=[m_1,\cdots,m_K]^T$, we may also express this relation
more compactly as $\PPu(\um)=\Pr\{\uu=\um\}$. The {\em
characteristic tensor} $\tbPPu\in\bbC^{P^{(\times K)}}$ is given
by the $K$-dimesional DFT of $\bPPu$, which, using a similar
index-vector notation, is given by
\begin{equation}
\tPPu(\un)=E[W_P^{\;\un^T\uu}]=\sum_{\um}\PPu(\um)W_P^{\;\un^T\um}.
\end{equation}
where the summation extends over all possible $P^K$ indices
combinations in $\um$.

\section{Problem Formulation and Indentifiability}
\label{sec:form}

We are now ready to formulate the mixture model over $\GFP$.
Assume that there are $K$ statistically independent random source
signals denoted $\us[t]=[s_1[t]\;s_2[t]\;\cdots\;s_K[t]]^T$, each
with an iid time-structure, such that at each time-instant $t$,
$s_k[t]$ is an independent realizations of a random variable in
$\GFP$, characterized by the (unknown) distribution vector
$\up_k$.

Let these sources be mixed (over $\GFP$) by an unknown, square
($K\times K$) mixing matrix $\A$ (with elements in $\GFP$),
\begin{equation}
\ux[t]=\A\circ\us[t].
\end{equation}

We further assume that $\A$ is invertible over the field, namely
that it has a unique inverse over $\GFP$, denoted
$\B\defeq\A^{-1}$, satisfying $\B\circ\A=\A\circ\B=\I$, where $\I$
denotes the $K\times K$ identity matrix. Like in ``classical"
linear algebra (over $\bbR$), $\A$ is non-singular (invertible)
iff its determinant\footnote{The determinant over $\GFP$ can be
calculated in a similar way to calculating the determinant over
$\bbR$, using the field's addition/subtraction and multiplication
operations.} is non-zero. Equivalently, $\A$ is singular iff there
exists (in $\GFP$) a nonzero vector $\uu$, such that $\A\circ
\uu=\uo$ (an all-zeros vector).

We are interested in the identifiability, possibly up to some
tolerable ambiguities, of $\A$ (or, equivalently, of its inverse
$\B$) from the set of observations $\ux[t]$, $t=1,2,...T$ under
asymptotic conditions, namely as $T\rightarrow\infty$. Due to the
assumption of iid samples for each source (implying ergodicity),
the joint statistics of the observations can be fully and
consistently estimated from the available data. Therefore, the
assumption of asymptotic conditions implies full and exact
knowledge of the joint probability distribution tensor $\bPPx$ of
the observation vector $\ux$ (we dropped the time-index $t$ here,
due to the stationarity). The remaining question is, therefore -
whether, and if so, under what conditions, $\A$ can be identified
(up to tolerable ambiguities) from exact, full knowledge of
$\bPPx$.

To answer this question, we first explore some basic statistical
properties of linear combinations of random variables over $\GFP$.
The characteristic vectors are particularly useful for this
analysis. Let $u$ and $v$ denote two statistically independent
random variables in $\GFP$ with probability vectors $\up_u$ and
$\up_v$ and characteristic vectors $\tup_u$ and $\tup_v$,
respectively. If $w=u\oplus v$, then the probability vector
$\up_w$ of $w$ is given by the cyclic convolution between $\up_u$
and $\up_v$, and the characteristic vector $\tup_w$ is therefore
given by the element-wise product of $\tup_u$ and $\tup_v$:
\begin{equation}
\label{conv}
p_w(n)=\sum_{m=0}^{P-1}\Pr\{u=m,v=n\ominus
m\}=\sum_{m=0}^{P-1}p_u(m)p_v(n\ominus
m)\;\;\Leftrightarrow\;\;\tp_w(n)=\tp_u(n)\tp_v(n)
\;\;\;\forall n.
\end{equation}

Two intuitively appealing (nearly trivial) properties follow from
this relation. First, combined with Property \ref{Pdeg} (in
Section \ref{sec:char}), this relation implies that the sum (over
$\GFP$) of two independent random variables is a degenerate random
variable iff both are degenerate. Likewise, combined with Property
\ref{Puni}, this relation implies that the sum is uniform if at least
one of the variables is uniform. The converse, however, is perhaps
somewhat less trivial, since it involves a distinction between
$\GFii$ and $\GFiii$ on one hand, and $\GFP$ with $P>3$ on the
other hand, as suggested by the following lemma:
\begin{lem}
\label{Lem1}
Let $u$ and $v$ be two statistically independent random variables
in $\GFP$, and let $w\defeq u\oplus v$. If both $u$ and $v$ are
non-uniform, then:
\begin{enumerate}
\item If $P=2$ or $P=3$, $w$ is
also non-uniform;
\item If $P>3$, $w$ may or may not be uniform.
\end{enumerate}
\end{lem}
\begin{proof}
By Property \ref{Puni}, $w$ would be uniform iff for each $n\ne
0$, either $\tp_u(n)=0$ or $\tp_v(n)=0$ (or both). In $\GFii$ this
can only happen if either $\theta_u=0$ or $\theta_v=0$ (or both),
which implies that at least one of the two variables is uniform.
Likewise, in $\GFiii$ this can only happen if either $\xi_u=0$ or
$\xi_v=0$ (or both), leading to a similar conclusion.

However, for $P>3$ there are sufficiently many degrees of freedom
in the characteristic vectors of $u$ and $v$ to allow both
non-zero and zero elements in both $\tup_u$ and $\tup_v$, as long
as at each $n\ne 0$ either one is zero. For example, consider
$P=5$ with $\tup_u=[1\;0\;0.3\;0.3\;0]^T$ and
$\tup_v=[1\;0.4\;0\;0\;0.4]^T$. This corresponds to
$\up_u\approx[0.32\;0.10\;0.24\;0.24\;0.10]^T$ and
$\up_v\approx[0.36\;0.25\;0.07\;0.07\;0.25]^T$, which are clearly
non-uniform. However, if these $u$ and $v$ are independent, their
sum (over $\mathbb{GF}(5)$) is a uniform random variable.
\end{proof}

Note, in addition, that since multiplication by a constant in
$\GFP$ is bijective, uniform or degenerate random variables cannot
become non-uniform or non-degenerate (nor vice-versa) by
multiplication with a constant. Consequently, the above
conclusions and Lemma \ref{Lem1} hold not only for the {\em sum}
of two random variables, but also for any {\em linear combination}
(over $\GFP$) thereof.

\noindent
We now add the following Lemma:
\begin{lem}
\label{Lem2}
Let $u$ and $v$ be two statistically independent, non-degenerate
random variables in $\GFP$, and let $w\defeq u\oplus v$. Then $v$
and $w$ are statistically independent iff $u$ is uniform.
\end{lem}
\begin{proof}
The joint probability distribution of $v$ and $w$ is given by
\begin{equation}
\label{Pvw}
P_{v,w}(m,n)=\Pr\{v=m,w=n\}=\Pr\{v=m,u=n\ominus
m\}=p_v(m)p_u(n\ominus m).
\end{equation}
Now, $w$ and $v$ are independent iff this probability equals
$p_v(m)p_w(n)$ for all $m,n$, namely iff $p_u(n\ominus m)=p_w(n)$
for all $n$ and for all $m$ with which $p_v(m)\ne 0$. Since $v$ is
non-degenerate, there are at least two such values of $m$.
Denoting these values as $m_1$ and $m_2$, this condition
translates into
\begin{equation}
p_u(n\ominus m_1)=p_u(n\ominus m_2)=p_w(n) \;\;\; \forall n.
\end{equation}
We therefore also have $p_u(n)=p_u(n\oplus m_1\ominus m_2)$
$\forall n$, which can be recursively generalized into
\begin{equation}
 p_u(n)=p_u(n\oplus
k\otimes(m_1\ominus m_2))\;\;\;\forall n,k\in\GFP.
\end{equation}
Since $P$ is prime, each element in $\GFP$ can be represented
(given $n$, $m_1$ and $m_2$) as $n\oplus k\otimes(m_1\ominus m_2)$
with some $k$, therefore this condition is satisfied iff $p_u(n)$
is constant, namely iff $u$ is uniform.
\end{proof}

To establish our identifiability condition we need one additional
lemma, which characterizes the entropy of a linear combination of
random variables in $\GFP$.
\begin{lem}
\label{Lem3}
Let $u$ and $v$ be two statistically independent, non-degenerate
random variables in $\GFP$, and let $w\defeq u\oplus v$. Then
$H(w)\ge H(u)$, where equality holds iff $u$ is uniform.
\end{lem}
\begin{proof}
As already mentioned in Section \ref{sec:char}, $H(w,v)\le
H(w)+H(v)$, with equality iff $w$ and $v$ are statistically
independent. In addition, $H(w|v)=H(w,v)-H(v)$. Therefore,
$H(w|v)\le H(w)$, with equality iff $w$ and $v$ are statistically
independent. Next, from \eqref{Pvw} we have
$\P_{w|v}(m,n)=p_u(n\ominus m)$, and therefore, as could be
intuitively expected,
\begin{equation}
H(w|v)=\sum_{m=0}^{P-1}p_v(m)\sum_{n=0}^{P-1}p_u(n\ominus m)\log
p_u(n\ominus m)=\sum_{m=0}^{P-1}p_v(m)H(u)=H(u),
\end{equation}
and we therefore conclude that $H(u)\le H(w)$, with equality iff
$w$ and $v$ are statistically independent. Now, according to Lemma
\ref{Lem2}, $w$ and $v$ are statistically independent iff $u$ is
uniform, which completes the proof.
\end{proof}
Obviously, a similar result (namely $H(w)\ge H(v)$) can be
obtained by switching roles between $u$ and $v$ in the proof. Note
an essential difference from a similar result over $\bbR$: In
$\bbR$ the entropy (or differential entropy) of a sum of two
independent, non-degenerate random variables is {\em always}
strictly larger than their individual entropies, no matter how
they are distributed. In $\GFP$, however, equality is attained if
one of the variables is uniform. In fact, this equality is
inevitable, simply because the entropy of any random variable in
$\GFP$ is upper-bounded by the uniform variable's entropy (of
$\log P$).

We are now ready to state our identifiability condition:
\begin{thm}
\label{Thm:id}
Let $\us$ be a $K\times 1$ random vector whose elements are
statistically-independent, non-degenerate random variables in
$\GFP$. Let $\A$ be a $K\times K$ non-singular matrix in $\GFP$,
and let the random vector $\ux$ be defined as $\ux=\A\circ\us$.
Assume that the probability distribution of $\ux$ is fully known
(specified by the probabilities tensor $\bPPx$). Then $\A$ can be
identified, up to possible permutation and scaling of its columns,
from $\bPPx$ alone, iff {\em none of the elements of $\us$ is a
uniform random variable}.
\end{thm}
\begin{proof}
The necessity of this condition is obvious by Lemma
\ref{Lem2}. Even in the simplest $2\times 2$ case, if one of the
sources, say $s_1$, is uniform, then by Lemma \ref{Lem2} any
linear combination of $s_1$ with the other source $s_2$ is still
statistically independent of $s_2$. Therefore, if the mixed
signals are $x_1=s_1\oplus s_2$ and $x_2=s_2$, then $x_1$ and
$x_2$ are statistically independent - so this situation is
indistinguishable from a non-mixing observation of two independent
sources with the same marginal distributions as $x_1$ and $x_2$
(which are also the marginal distributions of $s_1$ and $s_2$
(resp.) in this case).

To observe the sufficiency of the condition, note first that since
$\A$ is invertible over $\GFP$, any invertible linear mixture of
the original sources $\us$ can be obtained by applying some
invertible linear mixing to the observations $\ux$. Therefore, by
applying all (finite number of) invertible linear transformations
to $\ux$, one can implicitly obtain all the invertible linear
transformations of $\us$. Indeed, let $\hB$ denote an arbitrary
invertible matrix in $\GFP$, and denote
\begin{equation}
\uy\defeq\hB\circ\ux=(\hB\circ\A)\circ\us
\end{equation}
Since both $\hB$ and $\A$ are non-singular, so is $\hB\circ\A$,
which therefore:
\begin{enumerate}
\item  Has at least one non-zero element
in each row; and
\item Has at least one non-zero element
in each column, which means that each element of $\us$ is a
component of (namely, participates with nonzero weight in) at
least one element of $\uy$.
\end{enumerate}
Now define the respective sums of (marginal) entropies,
$\bH(\uy)\defeq\sum_{k=1}^KH(y_k)$ and
$\bH(\us)\defeq\sum_{k=1}^KH(s_k)$. Consequently, by Lemma
\ref{Lem3}, $\bH(\uy)$ cannot be made smaller than $\bH(\us)$.
Moreover, if none of the elements of $\us$ is uniform, then
\begin{equation}
\bH(\uy)=\bH(\us)\;\;\;\Leftrightarrow\;\;\;\hB\circ\A=\uPi\circ\uLam,
\end{equation}
where $\uPi$ denotes a $K\times K$ permutation matrix and $\uLam$
denotes a $K\times K$ diagonal, nonsingular matrix in $\GFP$. Any
other form of $\hB\circ\A$ would imply that at least one of the
elements of $\uy$ is a linear combination of at least two elements
of $\us$, and as such has higher entropy than both, and since at
least one of these two elements is also present in at least one
other element of $\uy$, $\bH(\uy)$ must be larger than $\bH(\us)$.

It is therefore possible, at least conceptually, to apply each
$K\times K$ nonsingular matrix $\hB$ in $\GFP$ to $\ux$, and
select one of the minimizers of $\bH(\uy)$. The inverse of this
minimizer is guaranteed to be equivalent to $\A$ up to permutation
and scaling,
\begin{equation}
\hB\circ\A=\uPi\circ\uLam\;\;\;\Leftrightarrow\;\;\;\hB^{-1}=\A\circ\uLam^{-1}\circ\uPi^T
\end{equation}
(where all the inverses are obviously taken over $\GFP$).
\end{proof}
Note that in $\GFii$ the scaling ambiguity is meaningless, because
the only possible scalar multiplication is by $1$, therefore only
the permutation ambiguity remains. In $\GFiii$ the possible
scaling ambiguity entails multiplication by either $1$ or $2$, or,
if the ``offset group" $\{0,1,-1\}$ is used, this ambiguity merely
translates into a sign-ambiguity.

Although the number of $K\times K$ nonsingular matrices in $\GFP$
is finite, this number is of the order of $P^{(K^2)}$, which
clearly becomes prohibitively large even with relatively small
values of $P$ and $K$. Therefore, our identifiability proof, which
is based on an exhaustive search, can hardly be translated into a
practical separation scheme. Nevertheless, in Section
\ref{sec:algos} below we shall propose and discuss two practical separation
approaches, which require a significantly reduced computational
effort. First, however, we need to address one more theoretical
aspect of our model - which is: whether (and if so under what
conditions) pairwise independence of linear mixtures implies their
full mutual independence.

\section{Pairwise independence implying full independence}
\label{sec:pair}

One of the basic, key concepts in ICA over $\bbR$ is the
Darmois-Skitovich Theorem (e.g., \cite{DST} p.218), which is used,
either explicitly or implicitly, in many ICA methods
(\cite{comon}). This theorem states that if two linear
combinations (over $\bbR$) of statistically independent random
variables are statistically independent, then all the random
variables which participate (with non-zero coefficients) in both
combinations must be Gaussian. Consequently (see, e.g.,
\cite{comon}), under the classical identifiability
condition (for ICA over $\bbR$) of not more than one Gaussian
source, pairwise statistical independence of linear mixtures of
the sources always implies their full mutual statistical
independence (namely, a non-mixing condition).

As we shall show in this section, this property does not carry
over to our $\GFP$ scenario by mere substitution of the Gaussian
distribution with the uniform. As it turns out, under our
identifiability condition (for ICA over $\GFP$) of no uniform
sources, pairwise independence implies full independence in
$\GFii$ and in $\GFiii$, but not in $\GFP$ with $P>3$. The reason
for this distinction is the distinction made in Lemma \ref{Lem1}
above, regarding the possibility that a linear combination of
non-uniform, independent random variables be uniform in $\GFP$
with $P>3$ (but not in $\GFii$ or in $\GFiii$).

Indeed, consider three independent random variables $s_1$, $s_2$
and $s_3$ in $\mathbb{GF}(5)$, with probability vectors
$\up_1=\up_2$ and $\up_3$ (resp.) following the example given in
the proof of Lemma
\ref{Lem1}. Namely, let the respective characteristic vectors be
given by $\tup_1=\tup_2=[1\;0\;0.3\;0.3\;0]^T$ and
$\tup_3=[1\;0.4\;0\;0\;0.4]^T$. This implies
$\up_1=\up_2\approx[0.32\;0.10\;0.24\;0.24\;0.10]^T$ and
$\up_3\approx[0.36\;0.25\;0.07\;0.07\;0.25]^T$. Clearly, our
identifiability condition is satisfied here, since none of these
random variables is uniform. However, $s_1\oplus s_3$, as well as
$s_2\oplus s_3$, are uniform. Thus, consider the mixing-matrix
$\A=\left[\begin{smallmatrix}1 & 0 & 0\\0 & 1 & 0\\1 & 1 &
1\end{smallmatrix}\right]$ which yields
\begin{equation}
\begin{split}
x_1&=s_1\\
x_2&=s_2\\
x_3&=s_1\oplus s_2\oplus s_3.
\end{split}
\end{equation}
Now, $x_1$ and $x_2$ are obviously statistically independent.
Moreover, since $s_2\oplus s_3$ is uniform and independent of
$s_1$, we deduce, by Lemma \ref{Lem2}, that $x_3$ and $x_1$ are
also statistically independent. Similarly, by switching roles
between $s_1$ and $s_2$, we further deduce that $x_3$ and $x_2$
are statistically independent as well. Therefore, $x_1$, $x_2$ and
$x_3$ are pair-wise independent, but are clearly not fully
mutually independent.

Obviously, such a counter-example cannot be constructed in $\GFii$
or in $\GFiii$, since in these fields a linear combination of
non-uniform, statistically independent random variables cannot be
uniform. Furthermore, out following theorem asserts that, under
our identifiability conditions, pairwise statistical independence
of the mixtures indeed implies their full statistical independence
in $\GFii$ and in $\GFiii$.

\begin{thm}
\label{Thm:pairs}
Let $\us$ be a $K\times 1$ random vector whose elements are
statistically-independent, non-degenerate and non-uniform random
variables in $\GFii$ or in $\GFiii$. Let $\uy=\D\circ\us$ denote a
$K\times 1$ vector of non-trivial linear combinations of the
elements of $\us$ over the field, prescribed by the elements of
the $K\times K$ matrix $\D$.

If the elements of $\uy$ are all pairwise statistically
independent (namely, if $y_k$ is statistically independent of
$y_{\ell}$ for all $k\ne\ell$, $k,\ell\in\{1,\ldots K\}$), then
$\D=\uPi\circ\L$, where $\uPi$ is a $K\times K$ permutation matrix
and $\L$ is a $K\times K$ non-singular diagonal matrix in the
field. In other words, the elements of $\uy$ are merely a
permutation of the (possibly scaled) elements of $\us$, and are
therefore not only pairwise, but also fully statistically
independent.
\end{thm}

Obviously, in $\GFii$ $\L$ must be $\I$ (no scaling ambiguity),
and in $\GFiii$ (assuming the group $\{0,1,-1\}$), $\L$ has only
$\pm 1$-s along its diagonal (the scaling ambiguity is just a sign
ambiguity). A proof for each of the two cases, $\GFii$ and
$\GFiii$, is provided in Appendix A. We now proceed to propose
practical separation approaches.

\section{Practical Separation Approaches}
\label{sec:algos}

In this section we propose two possible practical separation
approaches, based on the properties developed above.

Note that any approach which exploits the full statistical
description of the joint probability distribution of $\ux$ would
require collection (estimation) and some manipulation of the
probabilities tensor $\bPPx$, which is $P^K$ large, and,
therefore, a computational load of at least $\mathcal{O}(P^K)$
seems inevitable. Still, this is significantly smaller (and often
realistically far more affordable) than
$\mathcal{O}(K^2P^{(K^2)})$ (as required by brute-force search for
the unmixing matrix), even for relatively small values of $P$ and
$K$.

Note further, that in order to obtain reasonable estimates of
$\bPPx$ in practice, the number of available observation vectors
$T$ has to be significantly larger than $P^K$ (the size of
$\bPPx$). The estimation of $\bPPx$ can be obtained by the
following simple collection process:
\begin{enumerate}
\item Initialize $\hbPPx$ as an all-zeros tensor;
\item For $t=1,2,...,T$, set
$\hbPPx(\ux[t])\leftarrow\hbPPx(\ux[t])+1$;
\item Set $\hbPPx\leftarrow\tfrac{1}{T}\cdot\hbPPx$.
\end{enumerate}
Fortunately, however, a single collection of the observation's
statistics for obtaining $\hbPPx$ is generally sufficient, since,
in order to obtain the empirical statistical characterization
$\hbPPy$ of any linear transformation $\uy=\G\circ\ux$ of the
observations (where $\G$ is an arbitrary $L\times K$ matrix with
elements in $\GFP$), it is not necessary to actually apply the
transformation to the $T$ available observation vectors and then
recollect the probabilities. The same result can be obtained
directly (without re-involving the observations), simply by
applying a similar accumulation procedure to the $K$-way tensor
$\hbPPx$ in constructing the $L$-way tensor $\hbPPy$:
\begin{enumerate}
\item Initialize $\hbPPy$ as an all-zeros tensor;
\item Running over all $P^K$ index-vectors $\ui$ (from $[0\;\cdots\;0]^T$ to
$[P-1\;\cdots\;P-1]^T$), set
\begin{equation}
\label{updP}
\hbPPy(\G\circ\ui)\leftarrow\hbPPy(\G\circ\ui)+\hbPPx(\ui).
\end{equation}
\end{enumerate}
Note that when $\G$ is a square invertible matrix, $\hbPPy$ is
simply a permutation of $\hbPPx$.

\subsection{Ascending Minimization of EntRopies for ICA (AMERICA)}
Our first approach is based on minimizing the individual entropies
of the recovered sources. Conceptually, such an approach can
consist of going over all possible $P^K-1$ nontrivial linear
combinations of the observations, and computing their respective
entropies. Then, given these entropies, we need to select the $K$
linear combinations with the smallest entropies, such that their
respective linear-combination coefficients vectors (rows of the
implied unmixing matrix) are linearly independent (in $\GFP$).

Let us first consider the computation of the entropies of all
possible (nontrivial) $P^K-1$ linear combinations prescribed by
the coefficients vectors $\ui_n$ (for $n=1,...,P^K-1$). Each
requires the computation of the respective probabilities vector
$\up_{y_n}$ of $y_n=\ui_n^T\circ\ux$, by applying the
above-mentioned tensor-accumulation procedure with $\G=\ui_n^T$ to
the tensor $\hbPPx$. Thus, the number of required multiplications
is roughly $\mathcal{O}(K\cdot(P^K)^2)=\mathcal{O}(K\cdot
P^{2K})$, which (for $K>2$) is much smaller than
$\mathcal{O}(K^2\cdot P^{(K^2)})$ (the brute-force search cost),
but may still be quite large. Fortunately, it is possible to
compute the required probabilities vectors more conveniently, via
the estimated characteristic tensor $\htbPPx$, which can be
obtained using a multidimensional Fast Fourier Transform (FFT).

The proposed computation proceeds as follows. First, given the
estimated probabilities tensor $\hbPPx$, we obtain the estimated
characteristic tensor $\htbPPx$ using a $K$-dimensional FFT, by
successively applying $1$-dimensional radix-$P$ DFTs along each of
the $K$ dimensions. Thus, for each dimension we compute $P^{K-1}$
$P$-long DFTs, at the cost of $\mathcal{O}(P^{K-1}\cdot (P\log
P))=\mathcal{O}(P^K\log P)$. The total cost for obtaining
$\htbPPx$ is therefore $\mathcal{O}(K\cdot P^K\log
P)=\mathcal{O}(P^K\log(P^K))$, rather than $\mathcal{O}((P^K)^2)$,
as would be required by direct calculation.

Now, in order to obtain the characteristic vector $\tup_{y_n}$ of
$y_n=\ui_n^T\circ\ux$, we can exploit the following relation:
\begin{equation}
\tp_{y_n}(m)=E[W_P^{my_n}]=E[W_P^{m\ui_n^T\ux}]=\tPPx(m\otimes\ui_n),
\;\;\;m=0,...,P-1,
\end{equation}
which means that for each $\ui_n$, each ($m$-th) element of the
characteristic vector of $y_n$ can be extracted from the
respective element ($m\otimes\ui_n$) of $\tbPPx$. Note further,
that the first ($m=0$) element of each characteristic vector is
$1$; and that the conjugate-symmetry of the characteristic vectors
can be exploited, such that only the "first half"
($m=1,...,\lfloor P/2\rfloor$) needs to be extracted from
$\tbPPx$. Naturally, in the absence of the true $\tbPPx$, we would
use the empirical $\htbPPx$, obtained from the empirical
probabilities tensor $\hbPPx$, as described above.

The extraction of the characteristic vectors $\tup_{y_n}$ for all
$\ui_n$ requires $\mathcal{O}(P^K\cdot PK)$ additional operations.
Once these vectors are obtained, they are each converted, using
inverse FFT, into probabilities vectors $\up_{y_n}$, from which
the entropies are readily obtained. This requires additional
$\mathcal{O}(P^K\cdot(P\log P+P))$ operations (excluding the
computation of $P\cdot P^K$ logarithms).

Given the entropies of all possible linear combinations (ignoring
the trivial $\ui_0=\uo$), the one with the smallest entropy
corresponds to the first extracted source. Once the
smallest-entropy source is identified, a "natural" choice is to
proceed to the linear combination yielding the second-smallest
entropy (and so forth), but special care has to be taken, so that
each selected coefficients vectors should not be linearly
dependent (in $\GFP$) on the previous ones. One possible way to
assure this, is to take a ``deflation" approach (also sometimes
taken in classical ICA - see, e.g., \cite{deflation} or
\cite{ICA07}), in which each extracted source is first eliminated
from the mixture, and then the lowest-entropy combination of the
remaining (``deflated") mixtures is taken as the ``next" extracted
source. However, such an approach requires finding the
coefficients needed for elimination of the extracted source from
each mixture element, as well as recalculation of all the
entropies after each deflation stage, which seems computationally
expensive. A possible alternative is to use a greedy sequential
extraction, such that the $k$-th chosen coefficients vector is the
one associated with the smallest entropy while being linearly
independent of the previously selected $k-1$ coefficients vectors.
Checking whether a $K\times 1$ vector $\hub_k$ is linearly
independent of the $K\times 1$ vectors
$\hub_1,\hub_2,...,\hub_{k-1}$ amounts to checking whether there
exists a nonzero $k\times 1$ vector $\ual$, such that
$[\hub_1\;\cdots\;\hub_k]\circ\ual=\uo$, which can be checked by
an exhaustive search among all possible nonzero $k\times 1$
vectors in $\GFP$. This roughly adds (in the ``worst", last stage,
with $k=K$) $\mathcal{O}(K^2\cdot P^K)$ multiplications.

The total computational cost is therefore approximately
$\mathcal{O}(P^K\cdot(K^2+KP+K\log P + P\log P +P))$. The proposed
algorithm, which was given the acronym ``AMERICA" (Ascending
Minimization of EntRopies for ICA) is summarized in Table 1.

\vspace{0.5cm}
\noindent
\fbox{
\begin{minipage}[h]{\parwidth}
\fussy
\tt{
{\bf\underline{Algorithm 1: AMERICA}}\\
{\bf Input:} $\hbPPx$ - the mixtures' $K$-way $P\times
P\times\cdots\times P$ estimated (empirical)\\
probabilities tensor;\\
{\bf Output:} $\hB$ - the $K\times K$ estimated separation matrix;\\
{\bf Notations:} We denote by the $K\times 1$ $P$-nary vector
$\ui_n$ the $n$-th index\\ vector (for $n=0,...,P^K-1$), such that
$n=\sum_{k=1}^Ki_n(k)P^{k-1}$,\\ where
$\ui_n=[i_n(1)\;\cdots\;i_n(K)]^T$;
All indices in the description below run from $0$.\\
{\bf Algorithm:}
\begin{enumerate}
\item Compute $\htbPPx$, the observations' empirical characteristic
tensor,\\ by applying a $K$-dimensional radix-$P$ FFT to $\hbPPx$.
\item For $n=0,...,P^K-1$, compute $h_n$, the (empirical) entropy of
the\\ random variable $y_n\defeq \ui_n^T\circ\ux$ as follows:
\begin{enumerate}
\item Obtain the $P\times 1$ empirical characteristic vector
of $y_n$, denoted $\tup_n$, as follows:
\begin{enumerate}
\item Set $\tp_n(0):=1$;
\item Set $\tp_n(1):=\htPPx(\ui_n)$;
\item If $P=3$, set $\tp_n(2):=\htPPx^*(\ui_n)$;
\item If $P>3$, then for $m=2,...,(P-1)/2$, set
$\tp_n(m):=\htPPx(m\otimes\ui_n)$ and $\tp_n(P+1-m):=
\htPPx^*(m\otimes\ui_n)$;
\end{enumerate}
\item Obtain the $P\times1$ empirical probabilities vector of $y_n$,
denoted\\ $\up_n$, by applying an inverse FFT to the vector
$\tup_n$;
\item Obtain $h_n=\sum_{m=0}^{P-1}p_n(m)\log p_n(m)$;
\end{enumerate}
\item Find the smallest entropy among $h_1,...,h_{P^K-1}$ and
denote the\\ minimizing index $n_1$ (i.e., $h_{n_1}=\min_{n\ne
0}h_n$);
\item Set $\huB:=\ui_{n_1}^T$ and mark $h_{n_1}$ as "used";
\item Repeat for $k=2,...,K$:
\begin{enumerate}
\item \label{funused} Find the smallest among all "unused" entropies; denote
the\\
minimizing index $n_k$;
\item Construct the test-matrix $\buB:=[\huB^T\;\ui_{n_k}]$;
\item Go over all nonzero length-$k$ index vectors $\uj_n$
($n=1,...,p^k-1$),\\ checking whether $\buB\circ\uj_n=\uo$ for
some $n$. If such $\uj_n$ is found, mark $h_{n_k}$ as "used" and
find the next smaller entropy (i.e., go to\\ step \ref{funused});
\item Set $\huB:=\buB^T$.
\end{enumerate}
\end{enumerate}

}
\end{minipage}}
\vspace{0.5cm}

\subsection{Minimizing Entropies by eXchanging In COuples (MEXICO)}

An alternative separation approach, which avoids prior calculation
of the entropies of all possible linear combinations, is to try to
find the separating transformation by successively minimizing the
entropies in couples (going over all couples combinations in each
``sweep"). More specifically, let $x_1$ and $x_2$ denote the first
two elements of the mixtures vector, and let $\P_{1,2}$ denote
their $P\times P$ joint probability matrix, which can be obtained
from the tensor $\bPPx$ by summing along all other dimensions:
\begin{equation}
\P_{1,2}(m,n)=\sum_{i_3,\ldots,i_K=0}^{P-1}
\PPx(m,n,i_3,\ldots,i_K)\;\;\;
m,n\in[0,P-1].
\end{equation}
Consider a random variable of the form
\begin{equation}
\bar{x}_1=x_1\oplus c\otimes x_2,
\end{equation}
where $c\in[1,P-1]$ is some constant. Let $\up_{\bar{x}_1}(c)$
denote the probabilities vector of $\bar{x}_1$. The $m$-th element
of this vector is given (depending on $c$) by
\begin{equation}
\label{pfromP}
p_{\bar{x}_1}(m;c)=\Pr\{x_1\oplus c\otimes
x_2=m\}=\sum_{n=0}^{P-1}\Pr\{x_1=n,c\otimes x_2=m\ominus
n\}=\sum_{n=0}^{P-1}\P_{1,2}(n,c^{-1}\otimes(m\ominus n))\},
\end{equation}
where $c^{-1}$ denotes the reciprocal of $c$ in $\GFP$, such that
$c\otimes c^{-1}=1$. The entropy of $\bar{x}_1$ is then given by
\begin{equation}
H(\bar{x}_1;c)=-\sum_{m=0}^{P-1}p_{\bar{x}_1}(m;c)\log
p_{\bar{x}_1}(m;c).
\end{equation}
COnsider the value $c_0$ of $c$ which minimizes $H(\bar{x}_1;c)$.
If the resulting entropy is smaller than the entropy of $x_1$,
then substitution of $x_1$ with $\bar{x}_1=x_1\oplus c_0\otimes
x_2$ in $\ux$ would be an invertible linear transformation which
reduces the sum of entropies of the elements of $\ux$.

Note ,in addition, that following this transformation the mutual
information
$I(\bar{x}_1,x_2)=H(\bar{x}_1)+H(x_2)-H(\bar{x}_1,x_2)$ will be
smaller than $I(x_1,x_2)$, because the joint entropies
$H(x_1,x_2)$ and $H(\bar{x}_1,x_2)$ are the same (since the
transformation is invertible). Therefore, this transformation also
makes these two elements ``more independent".

Thus, based on this basic operation, a separation approach can be
taken as follows. Let $\uy$ denote the random vector of ``demixed"
sources to be constructed by successive linear transformations of
$\ux$, and initialize $\uy=\ux$, along with its probabilities
tensor $\bPPy=\bPPx$. Proceed sequentially through all couples
$y_k$, $y_\ell$ in $\uy$: For each couple, compute the joint
probabilities matrix $\P_{k,\ell}$, and then look for the value of
$c$ which minimizes the entropy of $\bar{y}_k=y_k\oplus c\otimes
y_\ell$. If this entropy is smaller than that of $y_k$, replace
$y_k$ with $\bar{y}_k$, recording the implied linear
transformation as $\buy=\V(k,\ell;c)\circ\uy$, where
\begin{equation}
\V(k,\ell;c)\defeq\I+c\cdot\uE_{k,\ell},
\end{equation}
$\uE_{k,\ell}$ denoting a $K\times K$ all-zeros matrix with a $1$
at the $(k,\ell)$-th position. If the minimal entropy of
$\bar{y}_k$ is larger than that of $y_k$, no update takes place,
and the next couple is addressed.

Upon an update, $\buy$ serves as the new $\uy$. The probabilities
tensor $\bPPy$ is updated accordingly (this update is merely a
permutation, attainable using
\eqref{updP} with $\G=\V(k,\ell;c)$). The procedure is
repeated for each indices-couple $(k,\ell)$ (with $k\ne \ell$),
and we term a ``sweep" as a sequential pass over all possible
$K(K-1)$ combinations (note that there is no symmetry here,
namely, the couple $(\ell,k)$ is essentially different from
$(k,\ell)$). Sweeps are repeated sequentially, until a full seep
without a single update occurs, which terminates the process.

In practice, the algorithm is applied starting with the empirical
observations' probabilities tensor $\hbPPx$, and the accumulated
sequential left-product of the $\V(k,\ell;c)$ matrices yields the
estimated separating matrix. Since the sum of marginal entropies
of the elements of $\uy$ is bounded below and is guaranteed not to
increase (usually to decrease) in each sweep, and since the
algorithm stops upon encountering the first sweep without such a
decrease - such a stop is guaranteed to occur within a finite
number of sweeps.

Note, however, that in general there is no guarantee for
consistent separation using this algorithm, i.e., even if the true
probabilities tensor $\bPPx$ of the observations is known (and
used), the stopping point is generally not guaranteed to imply
separation. The rationale behind this algorithm is the hope that
such a ``pairwise separation" scheme would ultimately yield
pairwise independence, which, at least for $P=2$ and $P=3$, would
in turn imply full independence (hence separation), per Theorem
\ref{Thm:pairs} above. Strictly speaking, however, this algorithm
is not even guaranteed to yield pairwise separation. For example,
consider the $P=2$ case, with a mixing matrix
\begin{equation}
\A=\begin{bmatrix}0 & 1 & 1 & 1\\
1 & 1 & 0 & 0\\
1 & 0 & 1 & 0\\
1 & 0 & 0 & 1\\
\end{bmatrix},
\end{equation}
when all the sources have equal $p(1)$ (probability of taking the
value $1$). In this particular case, the number of $1$-s in a
linear combination of any two lines is greater or equal to the
number of $1$-s in each of the two lines. Therefore, there is no
pairwise linear combination which reduces the entropy of any of
the mixtures in this case. Therefore, the algorithm may stop short
of full separation when such a condition is encountered.

Nevertheless, such conditions are relatively rare, and, as we show
in simulation results in the following section, this algorithm is
quite successful. Its leading advantage over AMERICA is in its
reduced computational complexity when the unmixing matrix $\B$ is
sparse and $K>>P$.

Indeed, the computational complexity of this iterative algorithm
naturally depends on the number of required sweeps and on the
number of updates in each sweeps - which in turn depend strongly
on the true mixing matrix $\A$ (and, to some extent, also on
sources' realizations). Testing each couple $(k,\ell)$ requires
computation of the joint probabilities matrix $\P_{k,\ell}$ -
which requires $\mathcal{O}(P^K)$ additions (no multiplications
are needed). Then, looking for the optimal $c$ requires $P-1$
computations of the probabilities vector of the respective
$\bar{y}_k$ - a total of additional $\mathcal{O}(P^3)$ additions
(again, no multiplications are needed for this) and
$\mathcal{O}(P^2)$ log operations. If an update takes place,
recalculation of $\hbPPy$ is also needed, which is
$\mathcal{O}(P^K)$ (but, as mentioned above, this is merely a
permutation of the tensor).

Therefore, the first sweep requires
$\mathcal{O}(P^2(P^K+P^3))=\mathcal{O}(P^{K+2}+P^5))$ operations
and $\mathcal{O}(P^4)$ log operations, plus $\mathcal{O}(P^K)$ for
each update within the sweep. Naturally, a couple tested in one
sweep does not have to be tested in a subsequent sweep if no
substitution involving any of its members had occurred in the
former. Therefore, for subsequent sweeps the number of operations
can be significantly smaller, depending on the number of updates
occurring along the way - which is obviously data-dependent. The
number of required sweeps is also data dependent.

Thus, the computational complexity of this algorithm, assuming
$K>3$, can be roughly estimated at
$\mathcal{O}(P^K\cdot(N_dP^2))$, where $N_d$ denotes a
data-dependent constant, which can be very small (of the order of
$2-3$) when the true demixing matrix $\B$ is very sparse (only a
few sweeps with few updates are needed), but can be considerably
large when $\B$ is rather ``rich". Compared to the computational
complexity of AMERICA, we observe that, assuming $K>>P$, this
algorithm is preferable if $N_dP^2<K^2$.

The algorithm, which was given the acronym ``MEXICO" (Minimizing
Entropies by eXchanging In COuples) is summarized in Table 2.

\vspace{0.5cm}
\noindent
\fbox{
\begin{minipage}[h]{\parwidth}
\fussy
\tt{
{\bf\underline{Algorithm 2: MEXICO}}\\
{\bf Input:} $\hbPPx$ - the mixtures' $K$-way $P\times P\times\cdots\times P$ estimated (empirical)\\
probabilities tensor;\\
{\bf Output:} $\hB$ - the $K\times K$ estimated separation matrix;\\
{\bf Algorithm:}
\begin{enumerate}
\item Initialize: $\hB:=\I$. Conceptually, we denote
the "demixed" random vector $\uy\defeq\hB\circ\ux$, so set
$\hbPPy:=\hbPPx$;
\item Initialize: $\uh=[h_1\cdots h_K]^T$ with the empirical
entropies of the $K$\\
respective elements $y_k$ (each computed from the empirical\\
probabilities vector, which is obtained by summation over all\\
other ($\ne k$) dimensions in $\hbPPy$);
\item Initialize $\F$, as a $K\times K$ all-ones flags matrix: $F(k,\ell)=1$ means that the $(k,\ell)$-th couple
needs to be (re)tested;
\item Run a "sweep": Repeat for $k=1,\ldots,K$, for
$\ell=1,\ldots,K$, $\ell\ne k$\\
If $F(k,\ell)=1$ do the following:
\begin{enumerate}
\item Compute $\P_{k,\ell}$, the empirical joint probabilities matrix
of $y_k$\\ and $y_\ell$, by summation over all other dimensions
($\ne k,\ell$) in $\hbPPy$;
\item For $c=1,\ldots,P-1$, compute the elements of
$\up_{\bar{y}_k}(c)$, the\\ probabilities vector of
$\bar{y}_k=y_k\oplus c\otimes y_\ell$, in a way similar to
\eqref{pfromP},\\ yielding its entropy $H(\bar{y}_k;c)$;
\item Denote the minimum entropy as $H_0=H(\bar{y}_k;c_0)$ (with $c_0$
denoting the minimizing $c$);
\item If $H_0<h_k$ apply a substitution:
\begin{enumerate}
\item Set $\V=\I+c_0\cdot\uE_{k,\ell}$;
\item Update $\hB:=\V\circ\hB$;
\item Update the probabilities tensor using \eqref{updP} with
$\G=\V$;
\item Mark all couples involving $k$ as "need to be retested":\\
$\F(k,:):=1$, $\F(:,k):=1$;
\item Update $h_k:=H_0$;
\item (Conceptually: $\uy:=\V\circ\uy$);
\end{enumerate}
\item Mark the $(k,\ell)$-th element as "tested": $F(k,\ell)=0$, and
proceed;
\end{enumerate}
\item If $\F\ne\I$ (there are still couples to be (re)tested), run
another sweep; Else stop.
\end{enumerate}

}
\end{minipage}}
\vspace{0.5cm}

\section{Rudimentary performance analysis and simulation results}
\label{sec:analsim}

In this section we present a rudimentary analysis of the expected
performance of the proposed algorithms, in order to obtain an
estimate of the expected rate of success in separating the
sources, at least in some simple cases.

Let us first establish the concept of {\em equivariance}. In
classical ICA, an algorithm is called equivariant (see, e.g.,
\cite{ca_equi}) with respect to the mixing matrix $\A$, if its
performance does not depend on $\A$ (as long as it is invertible),
but only on the realization of the sources. This appealing
property is shared by many (but certainly not by all) classical
ICA algorithms (in the context of noiseless classical ICA).

We shall now show that, with some slight modification, the AMERICA
algorithm is equivariant. Recall that AMERICA is based on
computation of all the empirical probabilities vectors $\up_{y_n}$
of the random variables $y_n=\ui_n^T\circ\ux$ for all possible
index-combinations $\ui_n$, followed by sequential extraction of
the index-vectors $\ui_n$ corresponding to the smallest entropies
(while maintaining sequential mutual linear independence).
Although not directly calculated in this way in the algorithm, the
$\ell$-th element of $\up_{y_n}$ is evidently given by
\begin{equation}
p_{y_n}(\ell)=\hPr\{\ui_n^T\circ\ux=\ell\}=\frac{1}{T}\sum_{t=1}^TI\{\ui_n^T\circ\ux[t]=\ell\},
\end{equation}
where $\hPr\{\cdot\}$ denoted the empirical probability, and where
$I\{\cdot\}$ denotes the Indicator function. But since
$\ux[t]=\A\circ\us[t]$, we obviously have
\begin{equation}
I\{\ui_n^T\circ\ux[t]=\ell\}=I\{(\A^T\circ\ui_n)^T\circ\us[t]=\ell\},
\end{equation}
which means that with any given realization $\us[1],\cdots,\us[T]$
of the sources, the empirical probabilities vector $\up_{y_n}$ of
$y_n=\ui_n^T\circ\ux$ obtained when the mixing matrix is $\A$, is
equal to some empirical probabilities vector $\up_{y_m}$ of
$y_m=\ui_m^T\circ\ux$ obtained when the mixing matrix is $\I$
(i.e., when there is no mixing), such that $\ui_m=\A^T\ui_n$.
Since $\A$ is invertible, this relation is bijective, which
implies that the $P^K-1$ empirical probabilities vectors obtained
with any (invertible) mixing are merely a permutation of the same
set of $P^K-1$ vectors that would be obtained when the sources are
not mixed. Consequently, if, based on the empirical entropies of
these empirical vectors, the matrix
$\huB=[\ui_{n_1}\;\ui_{n_2}\;\cdots\;\ui_{n_K}]^T$ is formed by
the algorithm when the mixing-matrix is $\A$, this implies that
the matrix
\begin{equation}
\huB_0\defeq[\ui_{m_1}\;\ui_{m_2}\;\cdots\;\ui_{m_K}]^T=
\left(\A^T\circ[\ui_{n_1}\;\ui_{n_2}\;\cdots\;\ui_{n_K}]\right)^T=\huB\circ\A
\end{equation}
would be formed by the algorithm when the sources are unmixed.
Consequently, the overall mixing-unmixing matrix\footnote{This
matrix is sometimes also called the ``contamination" matrix,
describing the residual mixing (if any).} $\huB\circ\A$ in the
mixed case would equal the overall mixing-unmixing matrix
$\huB_0\circ\I=\huB\circ\A$ in the unmixed case. This means that,
no matter what the (invertible) mixing matrix is, the overall
mixing-unmixing matrix would be the same as would be obtained by
the AMERICA algorithm in the unmixed case - implying the desired
equivariance property.

There is, however, one small caveat that has to be considered. The
reasoning above assumes that the sequential progress of the
algorithm through the sorted empirical entropies for selecting,
testing (for linear dependence) and using the index-vectors is
uniquely determined by the calculated entropy values, and is
independent of the values of the index-vectors. This is generally
true, with one possible exception: If the set of empirical
entropies happens to contain a subset with equal entropies, the
(arbitrary) order in which the index-vectors within such a subset
are sorted is usually lexicographic - which introduces dependence
on the actual index values, and such dependence is not
permutation-invariant - thereby potentially introducing dependence
on the mixing matrix in turn. In order to avoid this condition,
any sub-group with equal empirical entropies should be somehow
inner-sorted in a way which is independent of corresponding
index-vectors values - e.g., by randomization. Note that the
occurrence of such a subset (with empirical entropies that are
exactly equal) becomes very rare when the number of observations
$T$ is large, but may certainly happen when $T$ is relatively
small. Note further, that with such randomization the attained
separation for a given realization depends not only on the
sources' realization, but also on this random sorting within
subsets (but not on the mixing matrix), and therefore only
statistical measures of the performance (e.g., the probability of
perfect separation) can be considered equivariant.

Having established the equivariance, we now proceed to analyze the
probability of perfect separation in the most simple case: $P=2$,
$K=2$. Thanks to the equivariance property we may assume, without
loss of generality, that the mixing matrix is the identity matrix,
$\A=\I$. Let $p_{s_1}(1)=\rho_1$ (resp., $p_{s_2}(1)=\rho_2$)
denote the probability with which the first (resp., second) source
takes the value $1$. Due to the assumed non-mixing conditions
($\A=\I$), these are also the probabilities of the "mixtures"
$x_1$ and $x_2$. To characterize the empirical probabilities
tensor $\hbPPx$, let us denote by $N_{00}$, $N_{01}$, $N_{10}$ and
$N_{11}$ the number of occurrences of $\ux[t]=[0\;0]^T$,
$\ux[t]=[0\;1]^T$,  $\ux[t]=[1\;0]^T$ and  $\ux[t]=[1\;1]^T$
(resp.) within the observed sequence of length $T$. Thus, the
elements of the $2\times 2$ empirical probabilities tensor (matrix
in this case) are $\hPPx(m_1,m_2)=N_{m_1,m_2}/T$, for
$m_1,m_2\in\{0,1\}$.

The empirical probability $\hp_{x_1}(1)$ of $x_1$ taking the value
$1$ is given by $\hPPx(1,0)+\hPPx(1,1)=(N_{10}+N_{11})/T$. The
empirical probability $\hp_{x_1\oplus x_2}(1)$ of the random
variable $x_1\oplus x_2$ taking the value $1$ is given by
$\hPPx(1,0)+\hPPx(0,1)=(N_{10}+N_{01})/T$. An identification error
would occur if the entropy associated with the latter be smaller
than that associate with the former (because then the (wrong)
linear combination vector $\ui_3^T=[1\;1]$ would be preferred by
the algorithm over the (correct) linear combination vector
$\ui_1^T=[1\;0]$ as a row in $\huB$).

In the $P=2$ case, the entropy is monotonically decreasing in the
distance of $p(1)$ (or $p(0)$) from $\tfrac{1}{2}$. Assuming that
$T$ is ``sufficiently large", the empirical $\hp_{x_1}(1)$ would
be close to its true value $\rho_1$, and the empirical
$\hp_{x_1\oplus x_2}(1)$ would be close to its true value
$\rho_1(1-\rho_2)+\rho_2(1-\rho_1)=\rho_1+\rho_2-2\rho_1\rho_2$.
Assuming that $\rho_1,\rho_2<\tfrac{1}{2}$, both $\rho_1$ and
$\rho_1+\rho_2-2\rho_1\rho_2$ are smaller than $\tfrac{1}{2}$, and
we can therefore assume that so are the empirical $\hp_{x_1}(1)$
and $\hp_{x_1\oplus x_2}(1)$. Thus, the empirical entropy
associated with the linear combination $x_1\oplus x_2$ would be
smaller than that associated with $x_1$ if
\begin{equation}
\hp_{x_1\oplus x_2}(1)<\hp_{x_1}(1) \;\; \Leftrightarrow \;\;
\tfrac{1}{T}(N_{10}+N_{01})<\tfrac{1}{T}(N_{10}+N_{11}) \;\; \Leftrightarrow \;\;
N_{01}<N_{11}.
\end{equation}

We are therefore interested in the probability of the event
$\Xi1:\;N_{01}<N_{11}$. Let us denote by $N_2\defeq N_{01}+N_{11}$
the number of occurrences of $x_2[t]=1$ in $[1,T]$. The
probability of $\Xi1$ can then be expressed as follows:
\begin{multline}
\Pr\{\Xi1\}=\Pr\{N_{01}<N_{11}\}=\Pr\{N_{11}>\tfrac{1}{2}N_2\}=
\sum_{M=1}^T\Pr\{N_2=M\; \cap \;N_{11}>\tfrac{1}{2}M\}=\\
\sum_{M=1}^T\Pr\{N_2=M\}\Pr\{N_{11}>\tfrac{1}{2}M|N_2=M\}.
\end{multline}
Due to the statistical independence between the sources (and
therefore between $x_1$ and $x_2$), given that $N_2=M$, the random
variable $N_{11}$ is simply the number of occurrences of
$x_1[t]=1$ among $M$ independent trials - a Binomial random
variable with $M$ trials and probability $\rho_1$, which we shall
denote as $N_{1,M}\sim B(M,\rho_1)$. Thus,
\begin{multline}
\label{prxi}
\Pr\{\Xi1\}=\Pr\{N_{01}<N_{11}\}=\sum_{M=1}^T\Pr\{N_2=M\}\Pr\{N_{1,M}>\tfrac{1}{2}M\}=\\
\sum_{M=1}^T\binom{T}{M}\rho_2^M(1-\rho_2)^{T-M}\cdot
\sum_{N=\left\lfloor\tfrac{M}{2}\right\rfloor+1}^M\binom{M}{N}\rho_1^N(1-\rho_1)^{M-N}.
\end{multline}
The inner sum is the complementary cumulative distribution
function of the binomial distribution, which can also be expressed
using the {\em normalized incomplete beta function}\footnote{See,
e.g., Binomial Distribution from Wikipedia [online], available:\\
{\tt http://en.wikipedia.org/wiki/Binomial\_distribution}},
\begin{multline}
\Pr\{N_{1,M}>\tfrac{1}{2}M\}=1-\Pr\{N_{1,M}\le\tfrac{1}{2}M\}=\\
1-I_{1-\rho_1}(M-\left\lfloor\tfrac{1}{2}M\right\rfloor,\left\lfloor\tfrac{1}{2}M\right\rfloor+1)=
I_{\rho_1}(\left\lfloor\tfrac{1}{2}M\right\rfloor+1,\left\lceil\tfrac{1}{2}M\right\rceil),
\end{multline}
with
\begin{equation}
I_p(n,m)\defeq n\binom{n+m-1}{m-1}\int_0^p
t^{n-1}(1-t)^{m-1}dt=1-I_{1-p}(m,n).
\end{equation}
Note further (from \eqref{prxi}), that the probability of $\Xi1$
can be expressed as
\begin{equation}
\Pr\{\Xi1\}=E\left[I_{\rho_1}(\left\lfloor\tfrac{1}{2}N_2\right\rfloor+1,\left\lceil\tfrac{1}{2}N_2\right\rceil)\right]
\end{equation}
(where the expectation is taken with respect to $N_2$). When
$\rho_2\cdot T$ is "sufficiently large" this probability may be
approximated by substituting $N_2$ with its mean,
$E[N_2]=\rho_2\cdot T$,
\begin{equation}
\Pr\{\Xi1\}\approx I_{\rho_1}(\left\lfloor\tfrac{\rho_2}{2}T\right\rfloor+1,\left\lceil\tfrac{\rho_2}{2}
T\right\rceil).
\end{equation}

The event $\Xi1$ is just one possible component of an error event
in which the algorithm would prefer the (wrong) linear combination
vector $\ui_3^T=[1\;1]$ over the (correct) linear combination
vector $\ui_1^T=[1\;0]$. Such an error may also happen when the
empirical entropies of $x_1$ and of $x_1\oplus x_2$ are equal,
namely when $N_{01}=N_{11}$: assuming that the algorithm makes a
random decision in such cases (to ensure mean equivariance, as
discussed above), the probability of an error being caused by this
event (denoted $\Xi2$) would be $\tfrac{1}{2}\Pr\{\Xi2\}$.
Evidently,
\begin{multline}
\Pr\{\Xi2\}=\Pr\{N_{01}=N_{11}\}=\sum_{M=0}^T\Pr\{N_2=M\}\Pr\{N_{1,M}=\tfrac{1}{2}M\}=\\
\sum_{M'=0}^{\left\lfloor T/2 \right\rfloor}\tbinom{T}{2M'}\rho_2^{2M'}(1-\rho_2)^{T-2M'}\cdot
\tbinom{2M'}{M'}\rho_1^{M'}(1-\rho_1)^{M'}=
\sum_{M=0}^{\left\lfloor T/2
\right\rfloor}\tfrac{T!(1-\rho_2)^T}{(T-2M)!(M!)^2}\left(\tfrac{\rho_2^2\rho_1(1-\rho_1)}{(1-\rho_2)^2}\right)^{M}.
\end{multline}
Note that since the event $N_{1,M}=\tfrac{1}{2}M$ can only happen
for even values of $M$, an approximation using the mean with
respect to $N_2$ (as used for $\Pr\{\Xi1\}$ above) would be far
less accurate, and would therefore not be pursued.

Summarizing this part of the error analysis, the probability that
the algorithm would wrongly prefer $\ui_3^T=[1\;1]$ over
$\ui_1^T=[1\;0]$ as a row in $\huB$ can be approximated as
\begin{equation}
\label{Perr_diffp}
\Pr\{\Xi1\}+\tfrac{1}{2}\Pr\{\Xi2\}\approx
I_{\rho_1}(\left\lfloor\tfrac{\rho_2}{2}T\right\rfloor+1,\left\lceil\tfrac{\rho_2}{2}
T\right\rceil)+
\tfrac{1}{2}\cdot
\sum_{M=0}^{\left\lfloor T/2
\right\rfloor}\tfrac{T!(1-\rho_2)^T}{(T-2M)!(M!)^2}\left(\tfrac{\rho_2^2\rho_1(1-\rho_1)}{(1-\rho_2)^2}\right)^{M}.
\end{equation}

An error in the ``opposite" direction occurs when the algorithm
prefers $\ui_3^T=[1\;1]$ over $\ui_2^T=[0\;1]$ as a row in $\huB$.
The probability of this kind of error is evidently given by the
same expressions by swapping the roles of $\rho_1$ and $\rho_2$. A
failure of the algorithm is defined as the occurrence of either
one of the two errors. Although they are certainly not mutually
exclusive, we can still approximate (or at least provide an
approximate upper-bound for) the probability of occurrence of
either one, by the sum of probabilities of occurrence of each.
Assuming, for further simplicity of the exposition, that
$\rho_1=\rho_2=\rho$, the approximate probability of failure is
given by
\begin{equation}
\label{Perr2}
\Pr\{{\rm Failure}\}\approx
2\cdot
I_{\rho}(\left\lfloor\tfrac{\rho}{2}T\right\rfloor+1,\left\lceil\tfrac{\rho}{2}
T\right\rceil)+
\sum_{M=0}^{\left\lfloor T/2
\right\rfloor}\tfrac{T!(1-\rho)^T}{(T-2M)!(M!)^2}\left(\tfrac{\rho^3}{1-\rho}\right)^{M}.
\end{equation}
Recall that two assumptions are necessary for this approximation
to hold: i) that $\rho$ is sufficiently smaller than $0.5$; and
ii) that $\rho\cdot T$ is sufficiently large.

\begin{figure}[t]
\hspace{-0.3cm}
\includegraphics[width=170mm,height=110mm]{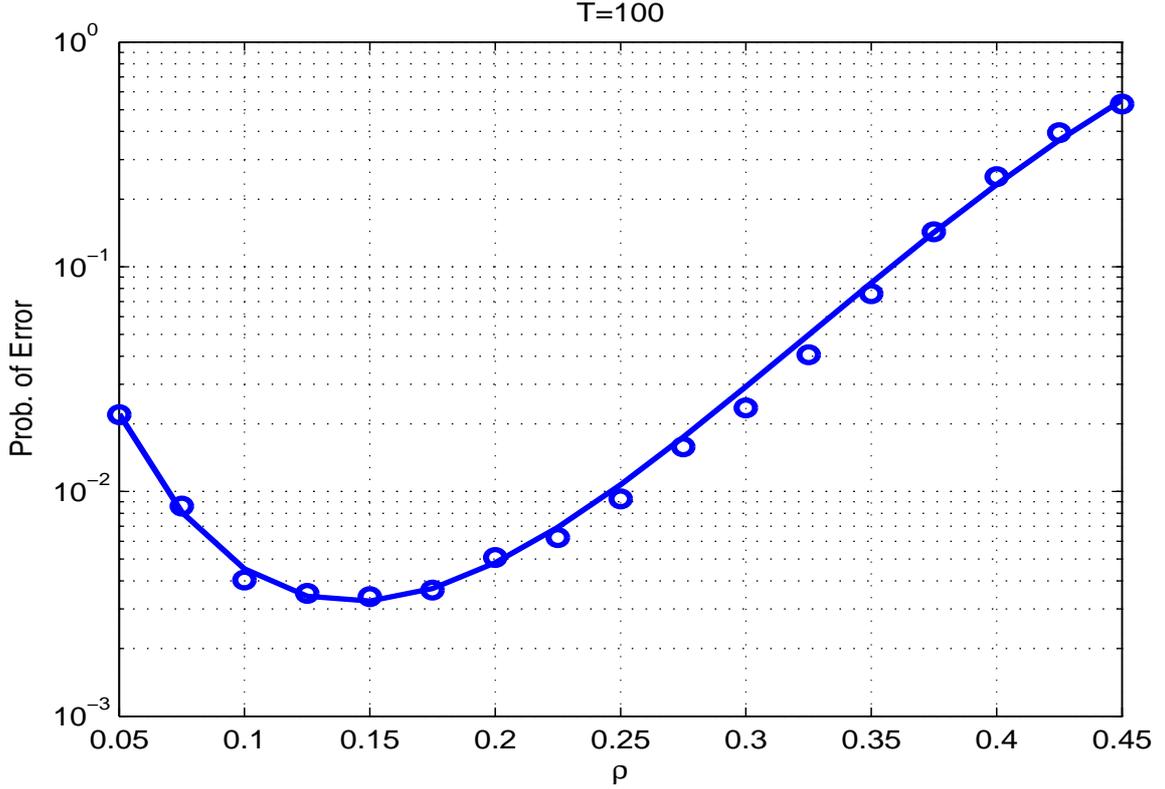}
\caption{\small Empirical probability of failure ('o') and its analytic approximation (solid) vs.\ the probability $\rho$ for
$P=2$, $K=2$ sources, $T=100$. The empirical probabilities were
obtained using $25,000$ independent trials}
\label{fig_Per_2x2}
\end{figure}

In order to test this approximation we simulated the mixing and
separation of $K=2$ independent binary ($P=2$) sources, each
taking the value $1$ with probability $\rho$. In Fig.\
\ref{fig_Per_2x2} we compare the analytic prediction \eqref{Perr2}
to the empirical probability of failure obtained in $25,000$
independent experiments (the sources and the mixing matrix were
drawn independently in each trial) vs. $\rho$ for $T=100$. Failure
of the separation is defined as the case in which $\huB\circ\A$ is
not a permutation matrix. We used the AMERICA algorithm for
separation (but for this ($K=2$) case, similar results are
obtained with MEXICO). The circles show the empirical
probabilities, whereas the solid line shows the approximate
analytic prediction
\eqref{Perr2}. The good match is evident.

When $K$ is larger than $2$, an approximate error expression can
be obtain by assuming that this type of error can occur
independently for each of the $K(K-1)/2$ different couples. Under
this approximate independence assumption, we get
\begin{equation}
\Pr\{{\rm Failure};K\}\approx1-(1-\Pr\{{\rm
Failure};K=2\})^{K(K-1)},
\end{equation}
where $\Pr\{{\rm Failure};K=2\}$ is given in \eqref{Perr2} above.
We assume here, for simplicity of the exposition, that all of the
sources take the value $1$ with similar probability $\rho$.
Extension to the case of different probabilities can be readily
obtained by using \eqref{Perr_diffp} for each (ordered) couple.

\begin{figure}[t]
\hspace{-0.3cm}
\includegraphics[width=170mm,height=110mm]{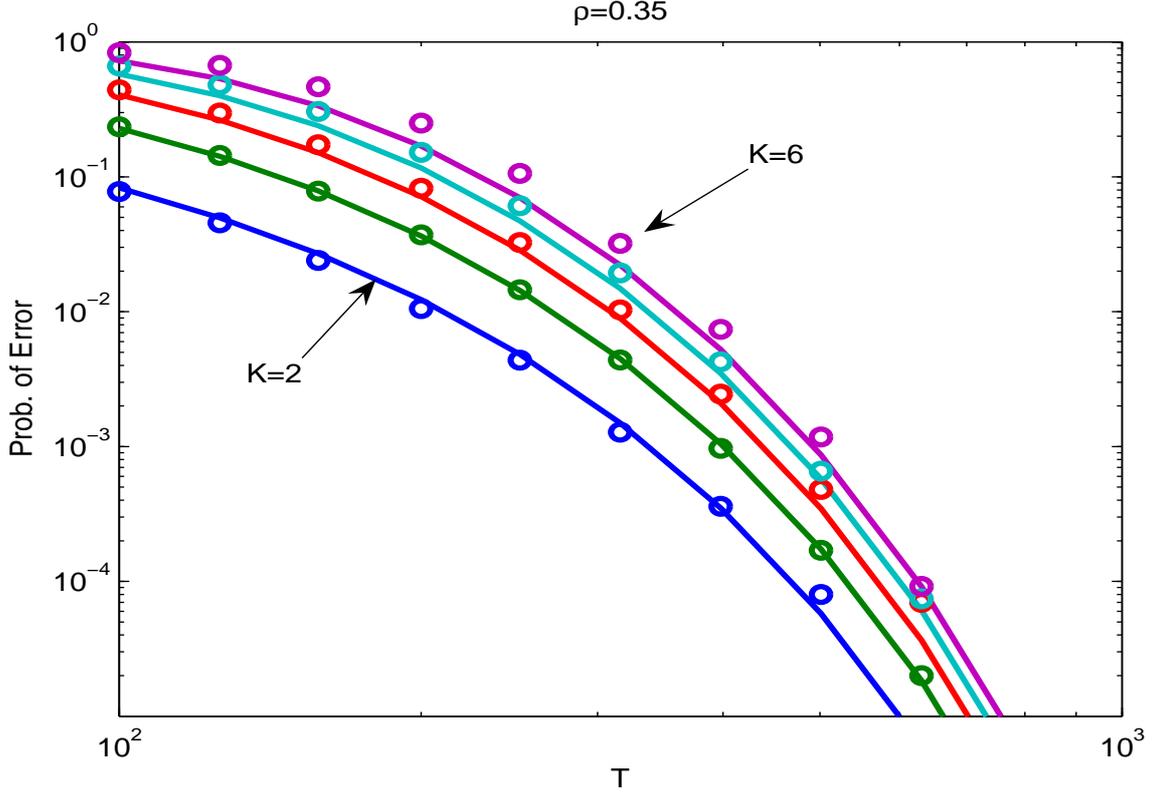}
\caption{\small Empirical probability of failure ('o') and its analytic approximation (solid) vs.\ the observation length $T$ for
$P=2$, $K=2,3,4,5$ and $6$ sources with $\rho=0.35$, using the
AMERICA algorithm. The empirical probabilities were obtained using
$100,000$ independent trials}
\label{fig_Perr_035_K2to6}
\end{figure}

To illustrate, we compare this expression in Fig.\
\ref{fig_Perr_035_K2to6} to the empirical probability of failure
(obtained in $100,000$ independent experiments) vs. $T$ for
$\rho=0.35$ with $K=2,3,4,5,6$. Again, failure of the separation
is defined as the case in which $\huB\circ\A$ is not a permutation
matrix (namely, any result which does not provide perfect
separation of {\em all} of the $K$ sources is considered a
``failure"). A good match is evident for the smaller values of
$K$, with some departure for the higher values - as could be
expected from the approximation induced by the error-independence
assumption.

\begin{figure}[t]
\hspace{-0.3cm}
\includegraphics[width=170mm,height=110mm]{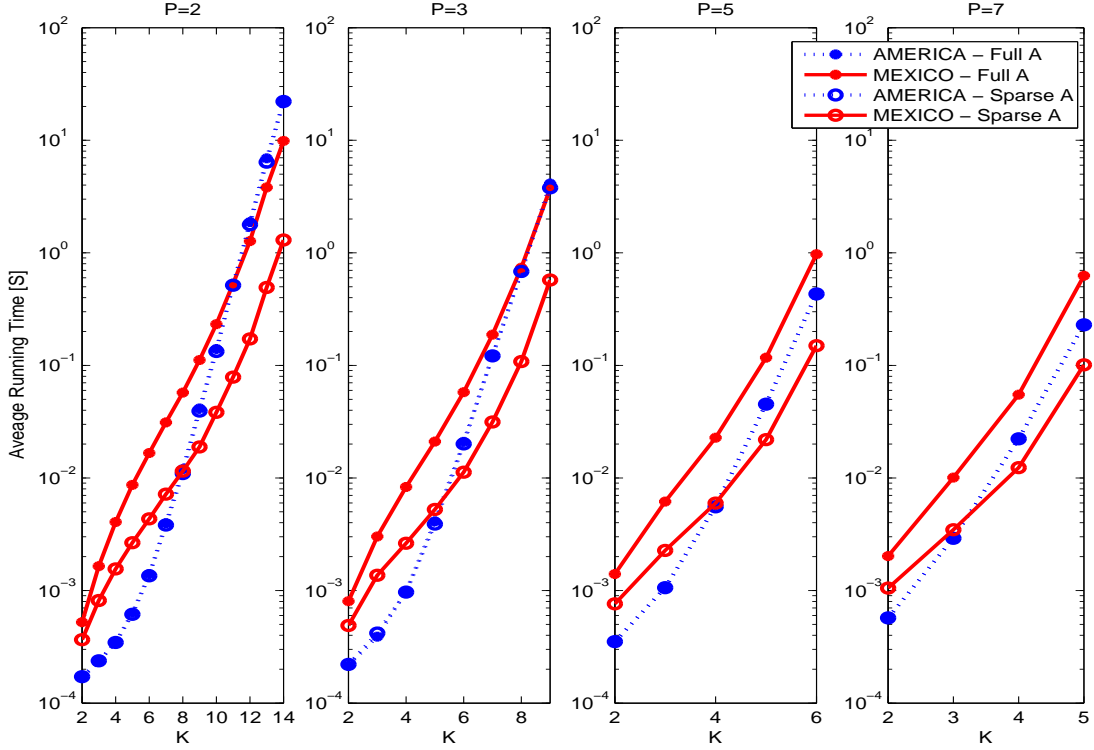}
\caption{\small Average running times (in [seconds]) for the AMERICA (dashed) and MEXICO (solid) algorithms, for full ('*') and sparse
('o') matrices. Note that the AMERICA plots for both the full and
the sparse mixing case are nearly identical.}
\label{fig_RunTime}
\end{figure}

Next, we compare the empirical, average running-times of the two
separation algorithm under asymptotic conditions. The
``asymptotic" conditions are emulated by substituting the
estimated (empirical) probabilities tensor $\hbPPx$ with the true
probabilities tensor $\bPPx$ as the input to the algorithms. We
simulated two cases: A ``full" mixing matrix and a ``sparse"
mixing matrix. The ``full" $K\times K$ (non-singular) mixing
matrices were randomly drawn in each trial as a product of a lower
triangular and an upper triangular matrix. The lower triangular
matrix $\uL$ was generated with random values independently and
uniformly distributed in $\GFP$ on and below the main diagonal,
substituting any $0$-s along the main diagonal with $1$-s; The
upper diagonal matrix $\uU$ was similarly generated by drawing all
values above the main diagonal, and setting the main diagonal to
all-$1$-s. Then $\A=\uU\circ\uL$. For generating the ``sparse"
matrices, the off-diagonal values of $\uL$ and $\uU$ were
``sparsified" by randomly (and independently) zeroing-out each
element, with probability $0.9$.

The elements of each of the sources' probabilities vectors
$\up_{s_1},\ldots\up_{s_K}$ were drawn uniformly in $(0,1)$ and
then normalized by their sum. The average running times (using
Matlab$^\circledR$ code \cite{MLcode} for both algorithms on a PC
Pentium$^\circledR$ 4 running at 3.4GHz) for several combinations
of $P$ and $K$ are shown in Fig. \ref{fig_RunTime}. Both
algorithms were applied to the same data, and the running times
were averaged over $4000$ independent trials. As expected, the
AMERICA algorithm is seen to be insensitive to the structure (full
/ sparse) of the mixing matrix; However, the MEXICO algorithm runs
considerably faster when $\A$ is sparse. Therefore, in terms of
running speed, MEXICO may be preferable when the mixing matrix is
known to be sparse, especially for relatively high values of $K$.

Note, however, that this advantage is somewhat overcast by a
degradation in the resulting separation performance. While perfect
separation was obtained (thanks to the ``asymptotic" conditions)
in all of the timing experiments by the AMERICA algorithm, few
cases of imperfect separation by MEXICO were encountered,
especially in the highest values of $K$ with the ``full" mixtures.

\begin{figure}[t]
\hspace{-0.3cm}
\includegraphics[width=170mm,height=110mm]{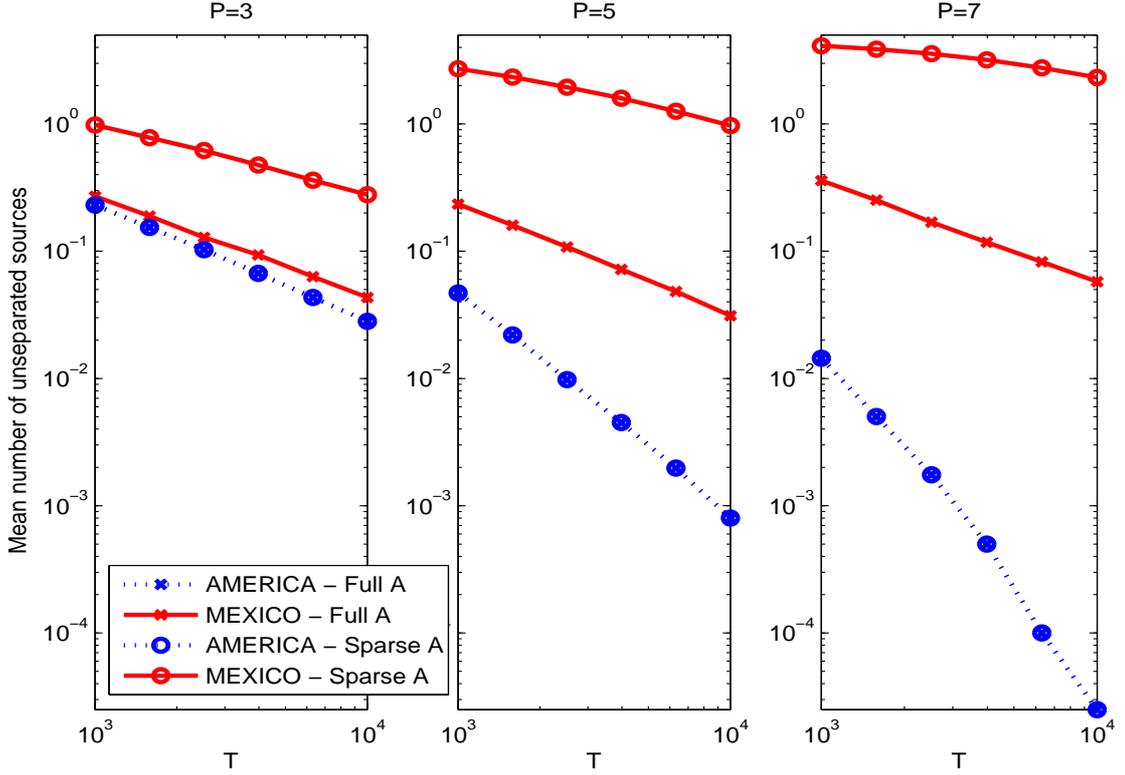}
\caption{\small Empirical mean number of unseparated sources (out
of the $K=5$ sources for AMERICA (dashed) and MEXICO (solid)
algorithms, for full ('*') and sparse ('o') matrices for
$P=3,5,7$. Each point reflect the average of $40,000$ trials. Note
that the AMERICA plots for both the full and the sparse mixing
case are nearly identical.}
\label{fig_RandP}
\end{figure}

To conclude this section, we provide (in Fig.\ref{fig_RandP}) some
empirical results showing the performance for $P=3,5,7$ with $K=5$
sources, with random sources' probabilities vectors. The
randomized elements of the probability vectors were independently
drawn (for each source, at each trial) from a uniform
distribution, and then normalized such that the sum of elements of
each probability vector adds up to $1$. The mixing matrix was
randomized at each trial as described above, once for a ``full
$\A$" and once for a ``sparse $\A$" version. In this experiment
the performance is measured as the mean number of unseparated
sources, which is defined (per trial) as the number of rows in the
resulting ``contamination matrix" $\huB\circ\A$ containing more
than one non-zero element (since, by construction of $\hB$ in both
MEXICO and AMERICA, $\huB\circ\A$ is always nonsingular, this is
exactly the number of sources which remain unseparated by the
algorithm). Each result on the plot reflects the average of
$40,000$ trials.

Evidently, the AMERICA algorithm seems significantly more
successful than the MEXICO algorithm, especially with the higher
values of $P$ (interestingly, the performance of AMERICA seems to
improve with the increase in $P$, whereas the performance of
MEXICO exhibits an opposite trend). The advantage of MEXICO is
confined to cases of small $P$ and large $K$, where its
potentially reduced computational load does not come at the
expense of a severe degradation in performance.

\section{conclusion}
\label{sec:conc}

We provided a study of general properties, identifiability
conditions and separation algorithms for ICA over Galois fields of
prime order $P$. We have shown that a linear mixture of
independent sources is identifiable (up to permutation and, for
$P>2$, up to scale) if and only if none of the sources is uniform.
We have shown that pairwise independence of an invertible linear
mixture of the sources implies their full independence (namely,
implies that the mixture is a scaled permutation) for $P=2$ and
for $P=3$, but not necessarily for $P>3$.

We proposed two different iterative separation algorithms: The
first algorithm, given the acronym AMERICA, is based on sequential
identification of the smallest-entropy linear combinations of the
mixtures. The second, given the acronym MEXICO, is based on
sequential reduction of the pairwise mutual information measures.
We provided a rudimentary performance analysis for $P=2$, which
applies to both algorithms with $K=2$, demonstrating a good fit of
the empirical results to the theoretical prediction. For higher
values of $K$ (still with $P=2$), we demonstrated a reasonable fir
up to $K\approx 6$ for the AMERICA algorithm.

AMERICA is guaranteed to provide consistent separation (i.e., to
recover all sources when the observation length $T$ is infinite),
and generally exhibits better performance (success rate) than
MEXICO with finite data lengths. However, when the mixing-matrix
is known to be sparse, MEXICO can have some advantage over AMERICA
is in its relative computational efficiency, especially for larger
values of $K$. Matlab$^\circledR$ code for both algorithms is
available online \cite{MLcode}.

Extensions of our results to common variants of the classical ICA
problem, such as ICA in the presence of additive noise, the
under-determined case (more sources than mixtures), possible
alternative sources of diversity (e.g., different temporal
structures) of the sources, etc. - are all possible. For example,
just like in classical ICA, temporal or spectral diversity would
enable to relax the identifiability condition, so as to
accommodate sources with uniform (marginal) distributions, which
might be more commonly encountered. However, these extensions fall
beyond the scope of our current work, whose main goal is to set
the basis for migrating ICA from the real- (or complex-) valued
algebraic fields to another.

\section*{Acknowledgement}
The author would like to thank Jacob Goldberger for inspiring
discussions on the use of FFT for the AMERICA algorithm.

\section*{Appendix A - A proof of Theorem 2}
In this Appendix we provide a proof of Theorem 2 for both $\GFii$
and $\GFiii$. Let $\us$ be a $K\times 1$ random vector whose
elements are statistically-independent, non-degenerate and
non-uniform random variables in either $\GFii$ or $\GFiii$, and
let $\uy=\D\circ\us$ denote a $K\times 1$ vector of non-trivial
linear combinations of the elements of $\us$ over the field,
prescribed by the elements of the $K\times K$ matrix $\D$ (in
either $\GFii$ or $\GFiii$, resp.).

Assume that $\D$ is a general matrix, and consider any pair $y_k$
and $y_{\ell}$ ($k\ne\ell$) in $\uy$. $y_k$ and $y_{\ell}$ are
linear combinations of respective groups of the sources, indexed
by the non-zero elements in $\D_{k,:}$ and $\D_{\ell,:}$, the
$k$-th and $\ell$-th rows (resp.) of $\D$.

Let us consider the case of $\GFii$ first.

\subsection{The $\GFii$ case}

The two groups composing $y_k$ and $y_{\ell}$ define, in turn,
three other subgroups (some of which may be empty):
\begin{enumerate}
\item Sub-group 1: Sources common to $\D_{k,:}$ and $\D_{\ell,:}$.
Denote the sum of these sources as $u$;
\item Sub-group 2: Sources included in $\D_{k,:}$ but excluded from $\D_{\ell,:}$.
Denote the sum of these sources as $v_1$;
\item Sub-group 3: Sources included in $\D_{\ell,:}$ but excluded from $\D_{k,:}$.
Denote the sum of these sources as $v_2$.
\end{enumerate}
For example, if (for $K=6$)
$\D_{k,:}=\left[\begin{array}{cccccc}0&1&1&1&1&1\end{array}\right]$
and
$\D_{\ell,:}=\left[\begin{array}{cccccc}1&1&0&0&1&1\end{array}\right]$,
then $u=s_2\oplus s_5\oplus s_6$, $v_1=s_3\oplus s_4$ and
$v_2=s_1$.

Note that by construction (and by independence of the elements of
$\us$), the random variables $u$, $v_1$ and $v_2$ are
statistically independent. Their respective probabilities vectors
and characteristic vectors are denoted
\begin{equation}
\up_\nu=\begin{bmatrix}p_\nu(0)\\ p_\nu(1)\end{bmatrix}\;,\;\;
\tup_\nu=\begin{bmatrix}1\\ \theta_\nu\end{bmatrix}\;,\;\;{\rm with}\;\;
\theta_\nu=1-2p_\nu(1)\;,\;\;{\rm for} \;\;\; \nu=u,v_1, v_2.
\end{equation}

Obviously, $y_k=u\oplus v_1$ and $y_{\ell}=u\oplus v_2$, so their
characteristic vectors are given by
\begin{equation}
\tup_{y_k}=\tup_{u}\odot\tup_{v_1}=\begin{bmatrix}
1\\\theta_{u}\theta_{v_1}\end{bmatrix}\;,\;\;
\tup_{y_\ell}=\tup_{u}\odot\tup_{v_2}=\begin{bmatrix}
1\\\theta_u\theta_{v_2}\end{bmatrix},
\end{equation}
where $\odot$ denotes the Hadamard (element-wise) product.

Define the random vector $\uw\defeq[y_k\;y_\ell]^T$, which can be
expressed as the sum of three independent random vectors:
\begin{equation}
\underbrace{\begin{bmatrix}y_k\\ y_\ell\end{bmatrix}}_{\uw}=
\underbrace{\begin{bmatrix}v_1\\ 0\end{bmatrix}}_{\defeq\uv_1}\oplus
\underbrace{\begin{bmatrix}u\\ u\end{bmatrix}}_{\defeq\uu}\oplus
\underbrace{\begin{bmatrix}0\\ v_2\end{bmatrix}}_{\defeq\uv_2}
\end{equation}
The probabilities matrices of the vectors $\uv_1$, $\uv_2$ and
$\uu$ are evidently given by
\begin{equation}
\P_{\uv_1}=\begin{bmatrix}1-p_{v_1}(1) & 0\\ p_{v_1}(1) &
0\end{bmatrix}\;\;\;
\P_{\uv_2}=\begin{bmatrix}1-p_{v_2}(1) & p_{v_2}(1)\\ 0 &
0\end{bmatrix}\;\;\;
\P_{\uu}=\begin{bmatrix}1-p_{u}(1) & 0\\0 &
p_u(1)\end{bmatrix}
\end{equation}
and therefore their characteristic matrices are given by
\begin{equation}
\tuP_{\uv_1}=\begin{bmatrix}1 & 1\\ \theta_{v_1} &
\theta_{v_1}\end{bmatrix}\;\;\;
\tuP_{\uv_2}=\begin{bmatrix}1 & \theta_{v_2}\\ 1 &
\theta_{v_2}\end{bmatrix}\;\;\;
\tuP_{\uu}=\begin{bmatrix}1 & \theta_u\\\theta_u &
1\end{bmatrix},
\end{equation}
where (see Section \ref{sec:char})
$\theta_\nu=E[W_2^\nu]=E[(-1)^\nu]=1-2p_\nu(1)$, for $\nu=v_1,
v_2, u$. Since $\uv_1$, $\uv_2$ and $\uu$ are statistically
independent, the characteristic matrix of $\uw$ is given by
\begin{equation}
\label{Pw1}
\tuP_{\uw}=\tuP_{\uv_1}\odot\tuP_{\uu}\odot\tuP_{\uv_1}=\begin{bmatrix}
1 & \theta_u\theta_{v_2} \\ \theta_{v_1}\theta_u &
\theta_{v_1}\theta_{v_2}\end{bmatrix}.
\end{equation}

On the other hand, if $y_k$ and $y_{\ell}$ are statistically
independent, the characteristic matrix of $\uw$ is also given by
\begin{equation}
\label{Pw2}
\tuP_{\uw}=\tup_{y_k}\tup_{y_{\ell}}^T=\begin{bmatrix}
1 & \theta_{v_2}\theta_u\\\theta_u\theta_{v_2} &
\theta_u^2\theta_{v_1}\theta_{v_2}\end{bmatrix}.
\end{equation}
Equating the expressions on \eqref{Pw1} and \eqref{Pw2}, we get
(only the $(2,2)$ element can differ)
\begin{equation}
\theta_u^2\theta_{v_1}\theta_{v_2}=\theta_{v_1}\theta_{v_2}.
\end{equation}
Since, due to Lemma \ref{Lem1}, if neither of the sources is
uniform, neither are $v_1$ and $v_2$, we have
$\theta_{v_1},\theta_{v_2}\ne0$, and therefore $\theta_u$ must be
either $1$ or $-1$. Since neither of the sources is degenerate,
this can only happen if $u=0$ (deterministically), which can only
happen if sub-group 1 is empty, namely, if the two rows $\D_{k,:}$
and $\D_{\ell,:}$ do not share common sources, or, in other words,
if there is no column $m$ in $\D$ such that both $\D_{k,m}$ and
$\D_{\ell,m}$ are $1$.

Applying this to all possible pairs of $k\ne\ell$ (for which $y_k$
and $y_\ell$ are independent), and recalling that $\D$ cannot have
any all-zeros row (no trivial combinations in $\uy$), we
immediately arrive at the conclusion that each row and each column
of $\D$ must contain exactly one $1$, meaning that $\D$ is a
permutation matrix.

We now turn to the case of $\GFiii$.

\subsection{The $\GFiii$ case}

For simplicity of the exposition, we shall now assume that the
values taken in $\GFiii$ are $\{0,1-1\}$ (rather than
$\{0,1,2\}$). Just like in the $\GFii$ case, we partition the two
groups composing $y_k$ and $y_{\ell}$ into subgroups, but now the
first (``common") subgroup is further partitioned into three
sub-subgroups:
\begin{enumerate}
\item Sub-group 1: Sources common to $\D_{k,:}$ and $\D_{\ell,:}$.
We partition this sub-group into four sub-subgroups according to
the coefficients in the respective rows of $\D$ as follows:
\begin{itemize}
\item Sub-subgroup 1a: sources for which the respective coefficients
in $\D_{k,:}$ and $\D_{\ell,:}$ are both $1$; Denote the sum of
these sources as $u_{++}$;
\item Sub-subgroup 1b: sources for which the respective coefficients
in $\D_{k,:}$ and $\D_{\ell,:}$ are both $-1$; Denote the sum of
these sources as $u_{--}$;
\item Sub-subgroup 1c: sources for which the respective coefficients
in $\D_{k,:}$ and $\D_{\ell,:}$ are $1$ and $-1$, resp.; Denote
the sum of these sources as $u_{+-}$;
\item Sub-subgroup 1d: sources for which the respective coefficients
in $\D_{k,:}$ and $\D_{\ell,:}$ are $-1$ and $1$, resp.; Denote
the sum of these sources as $u_{-+}$;
\end{itemize}
\item Sub-group 2: Sources included in $\D_{k,:}$ but excluded from $\D_{\ell,:}$.
Denote the respective linear combination of these sources as
$v_1$;
\item Sub-group 3: Sources included in $\D_{\ell,:}$ but excluded from $\D_{k,:}$.
Denote the respective linear combination of these sources as
$v_2$.
\end{enumerate}
For example, if (for $K=6$)
$\D_{k,:}=\left[\begin{array}{cccccc}0&1&-1&1&1&1\end{array}\right]$
and
$\D_{\ell,:}=\left[\begin{array}{cccccc}-1&-1&0&0&1&1\end{array}\right]$,
then $u_{++}=s_5\oplus s_6$, $u_{--}=u_{-+}=0$, $u_{+-}=s_2$,
$v_1=-s_3\oplus s_4$ and $v_2=-s_1$.

The random variables $u_{++}$, $u_{--}$, $u_{+-}$, $u_{-+}$, $v_1$
and $v_2$ are statistically independent. Their respective
probabilities vectors and characteristic vectors are denoted
\begin{equation}
\up_\nu=\begin{bmatrix}p_\nu(0)\\ p_\nu(1)\\ p_\nu(2)\end{bmatrix}\;,\;\;
\tup_\nu=\begin{bmatrix}1\\ \xi_\nu\\ \xi^*_\nu\end{bmatrix}\;,\;\;{\rm for} \;\;\;
\nu=u_{++}, u_{--}, u_{+-}, u_{-+}, v_1, v_2.
\end{equation}
An expression for $\xi_\nu=E[W_3^\nu]$ in terms of $p_\nu(0)$,
$p_\nu(1)$ and $p_\nu(2)$ can be found in \eqref{xiu} above. Note
further, that $\xi_{-\nu}=\xi_\nu^*$, so that
$\tup_{-\nu}=\tup_\nu^*$.

Evidently,
\begin{align}
y_k = v_1\oplus u_{++} \ominus u_{--} \oplus u_{+-} \ominus
u_{-+};\;,\;\; y_\ell = v_2\oplus u_{++} \ominus u_{--} \ominus
u_{+-}
\oplus u_{--},
\end{align}
so their characteristic vectors are given by
\begin{align}
\label{tupy}
\tup_{y_k}&=\tup_{v_1}\odot\tup_{u_{++}}\odot\tup_{u_{--}}^*\odot\tup_{u_{+-}}\odot\tup_{u_{-+}}^*
\nonumber
\\
\tup_{y_\ell}&=\tup_{v_2}\odot\tup_{u_{++}}\odot\tup_{u_{--}}^*\odot\tup_{u_{+-}}^*\odot\tup_{u_{-+}}
\end{align}

The random vector $\uw\defeq[y_k\;y_\ell]^T$ can now be expressed
as the sum of five independent random vectors:
\begin{equation}
\underbrace{\begin{bmatrix}y_k\\ y_\ell\end{bmatrix}}_{\uw}=
\underbrace{\begin{bmatrix}v_1\\ 0\end{bmatrix}}_{\defeq\uv_1}\oplus
\underbrace{\begin{bmatrix}u_{++}\\ u_{++}\end{bmatrix}}_{\defeq\uu_{++}}\oplus
\underbrace{\begin{bmatrix}-u_{--}\\ -u_{--}\end{bmatrix}}_{\defeq\uu_{--}}\oplus
\underbrace{\begin{bmatrix}u_{+-}\\ -u_{+-}\end{bmatrix}}_{\defeq\uu_{+-}}\oplus
\underbrace{\begin{bmatrix}-u_{-+}\\ u_{-+}\end{bmatrix}}_{\defeq\uu_{-+}}\oplus
\underbrace{\begin{bmatrix}0\\ v_2\end{bmatrix}}_{\defeq\uv_2}
\end{equation}
The probabilities matrices of the vectors $\uv_1$, $\uv_2$,
$\uu_{++}$, $\uu_{--}$, $\uu_{+-}$ and $\uu_{-+}$, and their
respective characteristic matrices are given by
\eqabcbegin
\begin{equation}
\P_{\uv_1}=\begin{bmatrix}p_{v_1}(0) & 0 & 0\\ p_{v_1}(1) &
0 & 0\\ p_{v_1}(2) & 0 & 0\end{bmatrix}\;\Rightarrow\;
\tuP_{\uv_1}=\begin{bmatrix}1 & 1 & 1\\ \xi_{v_1} &
 \xi_{v_1} &  \xi_{v_1}\\  \xi_{v_1}^* & \xi_{v_1}^* &
 \xi_{v_1}^*\end{bmatrix};
\end{equation}
\nexteqabc
\begin{equation}
\P_{\uv_2}=\begin{bmatrix}p_{v_2}(0) & p_{v_2}(1) & p_{v_2}(2) \\ 0 & 0 & 0\\
0 & 0 & 0\end{bmatrix}\;\Rightarrow\;
\tuP_{\uv_1}=\begin{bmatrix}1 & \xi_{v_2} & \xi_{v_2}^*\\
1 & \xi_{v_2} & \xi_{v_2}^*\\
1 & \xi_{v_2} & \xi_{v_2}^* \end{bmatrix};
\end{equation}
\eqabcend
\eqabcbegin
\begin{equation}
\P_{\uu_{++}}=\begin{bmatrix}
p_{u_{++}}(0) & 0 & 0\\
0 & p_{u_{++}}(1) & 0\\
0 & 0 & p_{u_{++}}(2)
\end{bmatrix}\;\Rightarrow\;
\tuP_{\uu_{++}}=\begin{bmatrix}
1 & \xi_{u_{++}} & \xi_{u_{++}}^*\\
\xi_{u_{++}} & \xi_{u_{++}}^* & 1\\
\xi_{u_{++}}^* & 1 & \xi_{u_{++}}\\
\end{bmatrix};
\end{equation}
\nexteqabc
\begin{equation}
\P_{\uu_{--}}=\begin{bmatrix}
p_{u_{--}}(0) & 0 & 0\\
0 & p_{u_{--}}(2) & 0\\
0 & 0 & p_{u_{--}}(1)
\end{bmatrix}\;\Rightarrow\;
\tuP_{\uu_{--}}=\begin{bmatrix}
1 & \xi_{u_{--}}^* & \xi_{u_{--}}\\
\xi_{u_{--}}^* & \xi_{u_{--}} & 1\\
\xi_{u_{--}} & 1 & \xi_{u_{--}}^*\\
\end{bmatrix};
\end{equation}
\nexteqabc
\begin{equation}
\P_{\uu_{+-}}=\begin{bmatrix}
p_{u_{+-}}(0) & 0 & 0\\
0 & 0 & p_{u_{+-}}(1)\\
0 & p_{u_{+-}}(2) & 0
\end{bmatrix}\;\Rightarrow\;
\tuP_{\uu_{+-}}=\begin{bmatrix}
1 & \xi_{u_{+-}}^* & \xi_{u_{+-}}\\
\xi_{u_{+-}} & 1 & \xi_{u_{+-}}^*\\
\xi_{u_{+-}}^* & \xi_{u_{+-}} & 1\\
\end{bmatrix};
\end{equation}
\nexteqabc
\begin{equation}
\P_{\uu_{-+}}=\begin{bmatrix}
p_{u_{-+}}(0) & 0 & 0\\
0 & 0 & p_{u_{-+}}(2)\\
0 & p_{u_{-+}}(1) & 0
\end{bmatrix}\;\Rightarrow\;
\tuP_{\uu_{+-}}=\begin{bmatrix}
1 & \xi_{u_{-+}} & \xi_{u_{-+}}^*\\
\xi_{u_{-+}}^* & 1 & \xi_{u_{-+}}\\
\xi_{u_{-+}} & \xi_{u_{-+}}^* & 1\\
\end{bmatrix};
\end{equation}
\eqabcend
Thus, the characteristic matrix of $\uw$ is given by the Hadamard
product of these matrices,
\begin{equation}
\tuP_{\uw}=\tuP_{v_1}\odot\tuP_{v_2}\odot\tuP_{u_{++}}\odot\tuP_{u_{--}}\odot\tuP_{u_{+-}}\odot\tuP_{u_{-+}}.
\end{equation}
Now, if $y_{k}$ and $y_\ell$ are statistically independent, then
$\tuP_\uw$ is also given by the outer product of their
characteristic vectors, which, using \eqref{tupy}, is given by
\begin{equation}
\tuP_\uw=\tup_{y_k}\tup_{y_\ell}^T=
(\tup_{v_1}\tup_{v_2}^T)\odot (\tup_{u_{++}}\tup_{u_{++}}^T)\odot
(\tup_{u_{--}}^*\tup_{u_{--}}^H)\odot
(\tup_{u_{+-}}\tup_{u_{+-}}^H)\odot
(\tup_{u_{-+}}^*\tup_{u_{-+}}^T),
\end{equation}
where $(\cdot)^H$ denotes the conjugate transpose. Noting that
$\tup_{v_1}\tup_{v_2}^T=\tuP_{v_1}\odot\tuP_{v_2}$, and recalling
that, since $v_1$ and $v_2$ cannot be uniform, $\xi_{v_1}$ and
$\xi_{v_2}$ must be non-zero, we conclude that the independence of
$y_k$ and $y_\ell$ implies that
\begin{equation}
(\tup_{u_{++}}\tup_{u_{++}}^T)\odot
(\tup_{u_{--}}^*\tup_{u_{--}}^H)\odot
(\tup_{u_{+-}}\tup_{u_{+-}}^H)\odot
(\tup_{u_{-+}}^*\tup_{u_{-+}}^T)=\tuP_{u_{++}}\odot
\tuP_{u_{--}}\odot\tuP_{u_{+-}}\odot\tuP_{u_{-+}}.
\end{equation}
It is easy to observe, that the first row and first column of each
of the matrices on the left-hand side (LHS) are indeed always
identical to those of the respective matrices on the right-hand
side (RHS), regardless of the values of the $\xi$ parameters. In
addition, in each of the matrices the $(2,2)$ element\footnote{In
this context (only) we enumerate these matrices' rows and columns
as $1,2,3$ rather than $0,1,2$.} is the conjugate of the $(3,3)$
element, and the $(2,3)$ element is the conjugate of the $(3,2)$
element. Therefore, the independence of $y_k$ and $y_\ell$ merely
implies the equality of the products of the $(2,2)$ elements on
the LHS and on the RHS, and of the products of the $(2,3)$
elements on the LHS and on the RHS.

The equality of the product of the $(2,2)$ elements implies
\eqabcbegin
\begin{equation}
\xi_{u_{++}}^*\cdot\xi_{u_{--}}\cdot1\cdot1=(\xi_{u_{++}})^2\cdot(\xi_{u_{--}}^*)^2
\cdot|\xi_{u_{+-}}|^2\cdot|\xi_{u_{-+}}|^2,
\end{equation}
\nexteqabc
and the equality of the product of the $(2,3)$ elements implies
\begin{equation}
1\cdot1\cdot\xi_{u_{+-}}^*\cdot\xi_{u_{-+}}=|\xi_{u_{++}}|^2\cdot|\xi_{u_{--}}|^2
\cdot(\xi_{u_{+-}})^2\cdot(\xi_{u_{-+}}^*)^2.
\end{equation}
\eqabcend
Taking the absolute values of both, and recalling that since
neither of the random variables $u_{++}$, $u_{--}$, $u_{+-}$ and
$u_{-+}$ can be uniform, neither of the $\xi$ parameters can be
zero, we have
\begin{align}
\label{all1}
|\xi_{u_{++}}|\cdot|\xi_{u_{--}}|&=|\xi_{u_{++}}|^2\cdot|\xi_{u_{--}}|^2
\cdot|\xi_{u_{+-}}|^2\cdot|\xi_{u_{-+}}|^2\;\Rightarrow\;
|\xi_{u_{++}}|\cdot|\xi_{u_{--}}|
\cdot|\xi_{u_{+-}}|^2\cdot|\xi_{u_{-+}}|^2=1\nonumber\\
|\xi_{u_{+-}}|\cdot|\xi_{u_{-+}}|&=|\xi_{u_{++}}|^2\cdot|\xi_{u_{--}}|^2
\cdot|\xi_{u_{+-}}|^2\cdot|\xi_{u_{-+}}|^2\;\Rightarrow\;
|\xi_{u_{++}}|^2\cdot|\xi_{u_{--}}|^2
\cdot|\xi_{u_{+-}}|\cdot|\xi_{u_{-+}}|=1.
\end{align}
Since for any random variable $\nu$ in $\GFiii$, $|\xi_\nu|\le 1$
with equality iff $\nu$ is degenerate, we conclude from
\eqref{all1} that if $y_k$ and $y_\ell$ are independent, then
$u_{++}$, $u_{--}$, $u_{+-}$ and $u_{-+}$ must all be degenerate.
Since none of the independent sources is degenerate, this implies,
in turn, that all four are identically zero, and that there are no
non-zero elements common to $\D_{k,:}$ and $\D_{\ell,:}$.

Like in the $\GFii$ case, by repeated application of this result
to all row-couples in $\D$, we conclude that pairwise independence
of the elements of $\uy$ implies that $\D$ is (up to signs)
permutation matrix, namely that the elements of $\uy$ are fully
mutually independent.

\bibliographystyle{ieee}
\bibliography{mexicoref}

%
\end{document}